\documentstyle[12pt]{article}

\newcommand{\al}{\alpha}
\newcommand{\ep}{\epsilon}

\newcommand{\la}{\lambda}

\newcommand{\deebar}{\bar{\partial}}

\newcommand{\df}{\stackrel{\rm def}{=}}

\newcommand{\lb}{\lbrack}
\newcommand{\rb}{\rbrack}

\newcommand{\msc}[1]{\mbox{\scriptsize #1}}
\newcommand{\dsp}{\displaystyle}

\newcommand{\nn}{\nonumber\\}

\newcommand{\br}{\mbox{{\bf R}}}
\newcommand{\bz}{\mbox{{\bf Z}}}

\newcommand{\bsz}{\msc{{\bf Z}}}

\newcommand{\id}{\mbox{Id}}

\newcommand{\cO}{{\cal O}}
\newcommand{\cN}{{\cal N}}
\newcommand{\cU}{{\cal U}}
\newcommand{\cM}{{\cal M}}

\newcommand{\ket}[1]{{\left|#1\right\rangle}}
\newcommand{\bra}[1]{{\left\langle#1\right|}}
\newcommand{\dket}[1]{{\left.\left|#1\right\rangle\right\rangle}}
\newcommand{\dbra}[1]{{\left\langle\left\langle#1\right|\right.}}

\newcommand{\Th}[2]{\Theta_{#1,#2}}
\newcommand{\th}{{\theta}}

\newcommand{\tr}{\mbox{Tr}}

\newcommand{\tJ}{\tilde{J}}
\newcommand{\tL}{\tilde{L}}
\newcommand{\tG}{\tilde{G}}
\newcommand{\tq}{\tilde{q}}
\newcommand{\tal}{\tilde{\al}}
\newcommand{\tX}{\tilde{X}}

\newcommand{\tpsi}{\tilde{\psi}}

\newcommand{\sNS}{\msc{NS}}

\newcommand{\sR}{\msc{R}}

\newcommand {\eqn}[1]{(\ref{#1})}

\makeatletter
\@addtoreset{equation}{section}
\def\theequation{\thesection.\arabic{equation}}
\makeatother

\begin{document}
\vskip 7mm

\begin{titlepage}
 \
 \renewcommand{\thefootnote}{\fnsymbol{footnote}}
 \font\csc=cmcsc10 scaled\magstep1
 {\baselineskip=14pt
 \rightline{
 \vbox{\hbox{hep-th/0307034}
       \hbox{UT-03-23}
       }}}

 \vfill
 \baselineskip=20pt
 \begin{center}
 \centerline{\Huge  Thermal Partition Functions for S-branes} 

 \vskip 2.0 truecm
\noindent{\it \large Yuji Sugawara} \\
{\sf sugawara@hep-th.phys.s.u-tokyo.ac.jp}
\bigskip

 \vskip .6 truecm
 {\baselineskip=15pt
 {\it Department of Physics,  Faculty of Science, \\
  University of Tokyo \\
  Hongo 7-3-1, Bunkyo-ku, Tokyo 113-0033, Japan}
 }
 \vskip .4 truecm

 \end{center}

 \vfill
 \vskip 0.5 truecm

\begin{abstract}
\baselineskip 6.7mm

We calculate the thermal partition functions of open strings 
on the S-brane backgrounds (the bouncing or rolling tachyon backgrounds)
both in the bosonic and superstring cases.
According to \cite{MSY}, we consider the discretized temperatures
compatible with the pure imaginary periodicity of tachyon profiles. 
The ``effective Hagedorn divergence'' is shown to appear no matter how low
temperature is chosen (including zero-temperature).
This feature is likely to be consistent with the large rate of 
open string pair production discussed in \cite{Strominger}
and also emission of closed string massive modes \cite{LLM}. 
We also discuss the possibility to remove the divergence 
by considering the space-like linear dilaton backgrounds as in \cite{KLMS}.

\end{abstract}

\vfill

\setcounter{footnote}{0}
\renewcommand{\thefootnote}{\arabic{footnote}}
\end{titlepage}
\baselineskip 18pt

\newpage
\section{Introduction}
\indent

Time dependent physics in string theory has been getting 
more and more  attentions 
and discussed by many theoretists from various angles.
The S(pace-like) brane \cite{GS1} has been providing 
an important laboratory to examine  the real time  dynamics of 
brane decay/creation processes caused by 
the open string tachyon condensation.   
Espicially,  the time-like boundary Sinh-Gordon type 
interaction \cite{Sen1,Sen2,Sen3}
$\dsp T(X^0) = \la \int_{\partial \Sigma} d\tau\, 
\cosh\left(X^0/\sqrt{\al'}\right)$ and the Liouville type interaction
\cite{Strominger,LNT}
$\dsp T(X^0) = \la \int_{\partial \Sigma} d\tau\, 
e^{\pm X^0/\sqrt{\al'}}$ have been intensively studied 
as the exactly soluble models of boundary conformal field theory 
(BCFT) describing the decay/creation processes of unstable D-branes. 
Following the terminology used in \cite{GS1,Strominger}, 
we shall call the Sinh-Gordon type model as the ``full S-brane''
and the Liouville type model as the ``half S-brane''.

Among other things, it has been discovered that 
these time dependent systems are followed by large rate of 
open string pair production \cite{Strominger,GS2} and 
closed string emission \cite{LLM}. It is an important fact 
that {\em all\/} the highly massive modes of string spectrum 
are excited by the decaying brane, which lead to
the Hagedorn like UV divergence and could destabilize the gas of 
``tachyon matter'' proposed in \cite{Sen1,Sen2,Sen3}
based on the classical analysis.  
The existence of such  Hagedorn like divergence  
has been already suggested 
in \cite{Strominger} and elucidated in \cite{MSY}. 
Related discussions about the particle creation 
and UV instability in the S-brane system 
are also found in the papers {\em e.g.\/} 
\cite{OY,Buchel,OS,CLL,Peet,Sen4,Gubser,DJ,Kluson,Sen6}. 

Another  remarkable  point  in the S-brane physics is the fact that 
we can define the thermal model {\em in spite of rapid time dependence,\/}
if assuming the discretized temperature compatible with the imaginary
periodicity of the boundary interaction \cite{MSY}.

Motivated by these studies, in this paper we calculate the thermal 
partition functions for the S-brane systems under the free string limit
$g_s\,\rightarrow\,0$, 
that is,  the Euclidean cylinder amplitudes 
with the thermal compactifications
compatible with the rolling (or bouncing) tachyon profiles.
We will further examine the thermodynamical behaviors of the 
partition functions. Although our conformal system is manifestly
positive definite because of the Euclidean signature, 
the modulus integral will be turnd out to show 
the Hagedorn like divergence {\em no matter how low
temperature is taken\/}.  This aspect seems to be consistent 
at least  qualitatively with 
the observations  given in \cite{Strominger,GS2,LLM} 
based on the caluculations in the Lorentzian signature. 
The recent works developing exact world-sheet analysis 
in the Lorentzian signature are 
given in \cite{GS2,ST,GIR,MV,KMS,CL,MTV,Schomerus,KLMS},
in particular, with emphasizing  interesting   relations  to the 
old (boundary) Liouville theory of two dimensional gravity and 
matrix models.

~

This paper is organized as follows.
Section 2 is the preliminary section.
We summarize some aspects of the $c=1$ ($c=3/2$)
boundary (super)conformal theory studied in  \cite{CKLM,PT,RS}.
Although nothing new is included in this section, 
we clarify some useful formulas to compute  the cylinder amplitudes.
In section 3, starting with presenting the thermal boundary states 
for the S-brane from the view points of affine $SU(2)$-current algebra,
we calculate the exact thermal partition function.
In section 4, we investigate the thermal behaviors of the partition 
functions, and find the strong thermal instabilities
irrespective of the temperature we choose. 
We also discuss the possibility to remove the UV divergence
in our Euclidean amplitudes by considering the space-like linear dilaton
backgrounds as in the recent paper \cite{KLMS}. 
Section 5 is devoted to presenting some discussions.

~

\section{Preliminary: Some Notes on Boundary Conformal Field Theory}
\indent

In this preliminary section we present a summary on the boundary
(super) conformal theory with $c=1$ ($c=3/2$) studied in
\cite{CKLM,PT,RS} for our later convenience.
Although nothing new is included in this section, 
we will clarify some useful relations
between the Virasoro Ishibashi states and those for 
the affine $SU(2)$-current algebra. The former is 
convenient to explicitly read off 
the interactions of S-branes with the closed string modes 
and has been treated in many literature
\cite{Sen1,Sen2,Sen3,Sen5,MSen,OS,LNT,ReyS}. 
On the other hand, the latter is 
useful for the calculation of cylinder amplitudes (especially, in 
the thermal model we are interested in), as will be turned out in 
the next section. 

~


\subsection{$c=1$ Boundary Conformal Field Theory}
\indent

We consider the free boson $X$ compactified on the circle of 
self-dual radius; $X\,\sim \, X + 2\pi $\footnote
   {We use the convention $\al'=1$ throughout  this paper. 
    Namely, we have the OPE; \\  
    $\dsp X_L(z)X_L(w) \sim -\frac{1}{2}\ln (z-w)$.}
As is well-known, we can realize the chiral $SU(2)_1$ current algebra 
in this set-up. Decomposing to the chiral sectors as 
$X(z,\bar{z})=X_L(z)+X_R(\bar{z})$, we define 
the $SU(2)_1$ currents as 
\begin{eqnarray}
&&J^3(z)=  i \partial X_L (z)~, ~~~ 
J^{\pm}(z) \left(\equiv J^1(z)\pm i J^2(z)\right) 
= e^{\pm i2 X_L(z)} ~, \nonumber \\
&&\tJ^3(\bar{z})= i \deebar X_R(\bar{z})~, ~~~ 
\tJ^{\pm}(\bar{z})\left(\equiv \tJ^1(\bar{z})\pm i \tJ^2(\bar{z})\right)
 = e^{\pm i 2X_R(\bar{z})}~.
\label{SU(2) 1 currents}
\end{eqnarray} 

The Neumann and Dirichlet boundary states are defined in the standard
manner
\begin{eqnarray}
&&\ket{N}= \frac{1}{2^{1/4}}\sum_{w\in\bsz}\,
\exp \left\{-\sum_{n=1}^{\infty}\frac{1}{n}\al_{-n}\tal_{-n}\right\} 
\ket{w}_L\otimes \ket{-w}_R~, 
\label{N}\\
&&\ket{D}= \frac{1}{2^{1/4}}\sum_{p\in\bsz}\,
\exp \left\{\sum_{n=1}^{\infty}\frac{1}{n}\al_{-n}\tal_{-n}\right\} 
\ket{p}_L\otimes \ket{p}_R~,
\label{D}
\end{eqnarray}
where $\al_n$, $\tal_n$ are the mode oscillators of $X(z)$, $\tX(\bar{z})$. 
The normalization factor $1/2^{1/4}$ is correctly chosen
by the standard Cardy condition \cite{Cardy}.

It is quite useful for our later analysis to introduce 
the Ishibashi states for the $SU(2)_1$ current algebra \cite{Ishibashi}. 
They are defined associated to each of the integrable representations 
$(\ell =0, 1)$, and characterized by the conditions 
\begin{eqnarray}
(J^a_n+\tJ^a_{-n})\dket{\ell}& =&0~, ~~~({}^{\forall} a, n) 
\label{glue SU(2)}
\\
\dbra{\ell} e^{-\pi s H^{(c)}} e^{2\pi i z J^3_0}
\dket{\ell'} &=&\delta_{\ell,\ell'} 
\chi^{(1)}_{\ell}(is, z) \nonumber \\
&\equiv & \delta_{\ell,\ell'} \frac{\Theta_{\ell,1}(is,z)}{\eta(is)}~,
\label{SU(2) Ishibashi}
\end{eqnarray}
where $\dsp H^{(c)}\equiv L_0+\tL_0-\frac{1}{12}$ is the closed string 
Hamiltonian, and $\chi^{(1)}_{\ell}(\tau, z)$ denotes the 
$SU(2)_1$ character of spin $\ell/2$ $(\ell =0,1)$.
It is also often convenient to consider the Ishibashi states for 
the Virasoro algebra associated to the primary states
of the degenerate representations:
$\ket{j,m,m'}$ ($\dsp j\in \frac{1}{2}\bz_{\geq 0}$), which compose  
the spin $j$ representation of the $SU(2)$ zero-mode algebras
$J_0^a$, $\tJ_0^a$; 
\begin{eqnarray}
&& L_0\ket{j,m,m'}=\tL_0\ket{j,m,m'}= j^2\ket{j,m,m'}~, \nonumber \\
&& L_n\ket{j,m,m'}=\tL_n\ket{j,m,m'}=0~,~~({}^{\forall} n >0) ~, \nonumber \\
&& J^3_0\ket{j,m,m'}=m\ket{j,m,m'}~,~~~\tJ^3_0\ket{j,m,m'}=m'\ket{j,m,m'}~.
\end{eqnarray} 
The corresponding Ishibashi states are defined by 
\begin{eqnarray}
(L_n-\tL_{-n})\dket{j,m,m'}&=&0~, ~~~ ({}^{\forall} n) ~, \nonumber \\
\dbra{j_1,m_1,m_1'}e^{-\pi s H^{(c)}} e^{2\pi i (z J^3_0+\bar{z}\tJ^3_0)} 
\dket{j_2,m_2,m_2'} &=& \delta_{j_1,j_2}\delta_{m_1,m_2}\delta_{m_1',m_2'} 
\, \chi^{\msc{Vir}}_{j}(is) e^{2\pi i z m_1}e^{2\pi i \bar{z}m_1'} \nonumber \\
&\equiv & \delta_{j_1,j_2}\delta_{m_1,m_2}\delta_{m_1',m_2'} \,
\frac{\tq^{j^2}-\tq^{(j+1)^2}}{\eta(is)}
e^{2\pi i z m_1}e^{2\pi i \bar{z}m_1'}~, \nonumber \\ 
\label{Virasoro Ishibashi}
\end{eqnarray}
where we wrote $\tq\equiv e^{-2\pi s}$ and $\chi^{\msc{Vir}}_j(is)$
denotes the Virasoro character of the degenerate representation with 
$h=j^2$.  The decomposition of the affine character
$\chi^{(1)}_{\ell}(\tau, z)$ by $\chi^{\msc{Vir}}_j(\tau)$ 
implies the obvious relations (under suitable choices of the phase factors
of Ishibashi states); 
\begin{eqnarray}
\dket{0}= \sum_{j\in \bsz_{\geq 0}}\,\sum_{m=-j}^j\, \dket{j,m,-m}~, ~~~
\dket{1}= \sum_{j\in \frac{1}{2}+\bsz_{\geq 0}}\,\sum_{m=-j}^j\, 
\dket{j,m,-m} ~.
\label{Ishibashi SU(2) Vir}
\end{eqnarray}
We can also easily show that 
\begin{eqnarray}
\ket{N}&=&\frac{1}{2^{1/4}}\left(\dket{0}+\dket{1}\right)~,
\label{N SU(2) Ishibashi} \\
e^{2\pi i J^3_0}\ket{N}&=&\frac{1}{2^{1/4}}\left(\dket{0}-\dket{1}\right)~,
\label{N SU(2) Ishibashi 2}
\end{eqnarray}
and $\ket{N}$, $e^{2\pi i J^3_0}\ket{N}$ compose the complete basis 
of $SU(2)_1$ Cardy states \cite{Cardy}
defined by the boundary condition \eqn{glue SU(2)}.

The Dirichlet boundary state \eqn{D} is similarly described 
by introducing the ``twisted'' Ishibashi states 
\begin{eqnarray}
\widehat{\dket{\ell}} \df e^{-i\pi J^1_0} \dket{\ell} \equiv 
   \sum_{j\in \frac{\ell}{2}+\bsz_{\geq 0}}\, e^{-i\pi j}
   \sum_{m=-j}^j \dket{j,m,m}~,
\label{twisted Ishibashi} 
\end{eqnarray}
which satisfies the boundary condition
\begin{eqnarray}
(J^3_n-\tJ^3_{-n})\widehat{\dket{\ell}}=0~,~~~ 
(J^{\pm}_n+\tJ^{\mp}_{-n})\widehat{\dket{\ell}}=0~,~~~({}^{\forall}n)~.
\label{twisted glue SU(2)}
\end{eqnarray}
We can then show 
\begin{eqnarray}
\ket{D} = e^{-i\pi J^1_0}\ket{N}= 
\frac{1}{2^{1/4}}\left(\widehat{\dket{0}}
   + \widehat{\dket{1}}\right)~,
\label{D SU(2) Ishibashi}
\end{eqnarray}
by a little calculation.
With the definition \eqn{twisted Ishibashi} we also obtain
\begin{eqnarray}
e^{i\pi J^1_0}\dket{0}= \widehat{\dket{0}}~,~~~ 
e^{i\pi J^1_0}\dket{1}= -\widehat{\dket{1}}~,
\end{eqnarray}
and hence 
\begin{eqnarray}
\ket{D'} \equiv 
e^{i\pi J^1_0}\ket{N} = \frac{1}{2^{1/4}}\left(\widehat{\dket{0}}
    -\widehat{\dket{1}}\right)~.
\label{D SU(2) Ishibashi 2}
\end{eqnarray}
\eqn{D SU(2) Ishibashi} and \eqn{D SU(2) Ishibashi 2} again compose 
the basis of $SU(2)_1$ Cardy states associated to the twisted boundary 
condition \eqn{twisted glue SU(2)}. 
The former corresponds to the periodic array of D-branes 
located at $X = 2\pi n $ ($n\in \bz$), while the latter
does to the configuration $X = 2\pi (n+1/2) $ ($n\in \bz$).
In fact, the twisted boundary condition \eqn{twisted glue SU(2)}
is equivalent with $J^3(z)=-\tJ^3(\bar{z})$, 
$J^{\pm}(z)=\tJ^{\mp}(\bar{z})$ in the open string channel, 
implying that $X\equiv X_L+X_R = 2\pi n$ or
$X\equiv X_L+X_R = 2\pi (n+1/2)$ $(n\in \bz)$.


~

\subsection{$c=3/2$ Boundary Superconformal Field Theory}
\indent

In order to generalize to the superstring case let us consider 
the $\cN=1$ boundary superconformal theory of $c=3/2$,
described by a system of one free boson and fermion $X$, $\psi$.
We assume the boson $X$ is compactified on a circle with 
the free fermion radius; $X\sim X+2\sqrt{2}\pi$.
The superconformal system is now described by three free fermions,
or equivalently,  the $SU(2)_2(\cong SO(3)_1)$ current algebra;
\begin{eqnarray}
&& J^3(z)=i\sqrt{2}\partial X_L(z)~, ~~~J^{\pm}(z)=\pm i \sqrt{2}\psi_L 
e^{\pm i \sqrt{2}X_L}(z)~,
\nonumber \\
&& 
\tJ^3(z)=i\sqrt{2}\deebar X_R(\bar{z})~, ~~~\tJ^{\pm}(z)= \pm i \sqrt{2}\psi_R 
e^{\pm i \sqrt{2}X_R}(\bar{z})~.
\label{SU(2) 2 currents} 
\end{eqnarray}
(The overall phase factors $\pm i$ in $J^{\pm}$ are chosen nothing but 
for the later convenience.)

The Neumann and Dirichlet boundary states are written as 
\begin{eqnarray}
&&\ket{N}= 
\exp \left\{-\sum_{n=1}^{\infty}\frac{1}{n}\al_{-n}\tal_{-n}
-i\sum_{r>0}\psi_{-r}\tpsi_{-r}\right\} 
\ket{N}^{(0)}~, 
\label{N super}\\
&&\ket{D}= 
\exp \left\{\sum_{n=1}^{\infty}\frac{1}{n}\al_{-n}\tal_{-n}
+i\sum_{r>0}\psi_{-r}\tpsi_{-r}\right\} 
\ket{D}^{(0)}~,
\label{D super}
\end{eqnarray}
where $r $ runs over $\dsp \frac{1}{2}+\bz$ ($\bz$) for NS (R) sector,
and $\ket{N}^{(0)}$, $\ket{D}^{(0)}$ denote the zero-mode parts
specified later\footnote
   {The convention of boundary states with the boundary conditions 
     $(\psi_r+i\ep\tpsi_{-r})\ket{N;\ep}=0$, \\
     $(\psi_r-i\ep\tpsi_{-r})\ket{D;\ep}=0$, ($\ep=\pm$) is 
     often used in literature, and the GSO projected boundary states
     are written as $\dsp \frac{1}{2}\left(\ket{B;+}-\ket{B;-}\right)$
     ($B=N$ or $D$). In this paper we shall fix $\ep=+$ and explicitly 
     insert the GSO projection operator when calculating  amplitudes.}.

As in the case of bosonic string it is useful to reexpress  
things by the language of $SU(2)$ current algebra.
We introduce the Ishibashi states $\dket{\ell}$ $(\ell =0,1,2)$ 
for $SU(2)_2$ as the ones satisfy the boundary conditions 
\eqn{glue SU(2)} and also 
\begin{eqnarray}
&& \dbra{\ell} e^{-\pi s H^{(c)}} e^{2\pi i z J^3_0}\dket{\ell'} 
= (-1)^{\ell}\delta_{\ell,\ell'} \chi^{(2)}_{\ell} (is,z) ~,  \\
&& \chi^{(2)}_0(\tau,z) \equiv
\frac{1}{2}\left(\sqrt{\frac{\th_3(\tau,0)}{\eta(\tau)}}
\frac{\th_3(\tau,z)}{\eta(\tau)}
+\sqrt{\frac{\th_4(\tau,0)}{\eta(\tau)}}
\frac{\th_4(\tau,z)}{\eta(\tau)}
\right) ~, \nonumber \\
&& 
 \chi^{(2)}_2(\tau,z) \equiv
\frac{1}{2}\left(\sqrt{\frac{\th_3(\tau,0)}{\eta(\tau)}}
\frac{\th_3(\tau,z)}{\eta(\tau)}
-\sqrt{\frac{\th_4(\tau,0)}{\eta(\tau)}}
\frac{\th_4(\tau,z)}{\eta(\tau)}
\right) ~, \nonumber \\
&&  \chi^{(2)}_1(\tau,z) \equiv
\frac{1}{\sqrt{2}}\sqrt{\frac{\th_2(\tau,0)}{\eta(\tau)}}
\frac{\th_2(\tau,z)}{\eta(\tau)}~,
\label{SU(2) 2 character}
\end{eqnarray} 
where $\chi^{(2)}_{\ell}(\tau,z)$ denotes the spin $\ell/2$ character of 
$SU(2)_2$. The overall factor $(-1)^{\ell}$ is  
the  phase convention that reproduces the correct $(-1)$ factor for
the space-time fermions.

We may also consider the Ishibashi states for the $\cN=1$ super Virasoro
algebra $\left\{L_n,\, G_r\right\}$
defined similarly to \eqn{Virasoro Ishibashi}.
The degenerate representations are again characterized by 
the spin $j$ of the $SU(2)$ zero-mode algebra and have conformal weights
$\dsp h=j^2/2$. To be more precise, we should restrict the spin $j$
as $j\in \bz_{\geq 0}$ for the NS sector and $\dsp j\in \bz_{\geq 0}
+ \frac{1}{2}$ for the R sector due to the locality of the 
$SU(2)$ currents \eqn{SU(2) 2 currents}. The corresponding Ishibashi
states are defined by
\begin{eqnarray}
 (L_n-\tL_{-n})\dket{j,m,m'}&=&0~, ~~~ ({}^{\forall} n) ~, \nonumber \\
 (G_r-i\tG_{-r}) \dket{j,m,m'}&=&0~, ~~~
\left\{
\begin{array}{ll}
  {}^{\forall} r \in \bz + \frac{1}{2} & ~\mbox{if } j\in \bz_{\geq 0} \\
  {}^{\forall} r \in \bz &~ \mbox{if } j\in \bz_{\geq 0} +\frac{1}{2}~.
\end{array}
\right.
 \nonumber \\
 \dbra{j_1,m_1,m_1'}e^{-\pi s H^{(c)}} e^{2\pi i (z J^3_0+\bar{z}\tJ^3_0)} 
\dket{j_2,m_2,m_2'} &=& \delta_{j_1,j_2}\delta_{m_1,m_2}\delta_{m_1',m_2'} 
\, \chi^{\msc{$\cN=1$}}_{j}(is) 
e^{2\pi i z m_1}e^{2\pi i \bar{z}m_1'} ~, \nonumber \\
\chi^{\cN=1}_j(\tau)&\equiv &  \frac{q^{\frac{j^2}{2}}-q^{\frac{(j+1)^2}{2}}}
{\eta(\tau)} \sqrt{\frac{\th_3}{\eta}}(\tau)~, ~~(j\in \bz_{\geq 0})
\nonumber \\ 
\chi^{\cN=1}_j(\tau)&\equiv &  \frac{q^{\frac{j^2}{2}}-q^{\frac{(j+1)^2}{2}}}
{\eta(\tau)} \sqrt{\frac{2\th_2}{\eta}}(\tau)~, ~~(j\in \frac{1}{2}+
\bz_{\geq 0})~, \nn
&&
\label{super Virasoro Ishibashi}
\end{eqnarray}
where $\chi^{\cN=1}_j(\tau)$ denotes the degenerate characters
of $c=3/2$ super Virasoro algebra of NS (R) sector 
for $j\in \bz_{\geq 0}$ ($\dsp j\in \frac{1}{2}+\bz_{\geq 0}$).
Similarly to \eqn{Ishibashi SU(2) Vir}
the decomposition of the affine character $\chi^{(2)}_{\ell}(\tau,z)$
by $\chi^{\cN=1}_j(\tau)$ leads to the relations
\begin{eqnarray}
 \dket{0}+\dket{2}&=& \sum_{j\in \bsz_{\geq 0}}\sum_{m=-j}^j\dket{j,m,-m}~,
\nonumber \\
 \dket{1} &=& \sum_{j\in \frac{1}{2}+\bsz_{\geq 0}}
\sum_{m=-j}^j\dket{j,m,-m}~.
\label{Ishibashi SU(2) sVir}
\end{eqnarray}
Moreover, the Neumann boundary state 
of NS sector \eqn{N super} is identified as 
\begin{eqnarray}
\ket{N}_{NS}=\dket{0}+\dket{2}~,
\label{N super NS}
\end{eqnarray}
with the standard zero-mode part
\begin{eqnarray}
\ket{N}_{NS}^{(0)}
= \sum_{w\in \bsz} \ket{\sqrt{2}w}_L\otimes \ket{-\sqrt{2}w}_R~.
\label{N NS 0}
\end{eqnarray} 
We shall also define for the R sector 
\begin{eqnarray}
\ket{N}_{R} = \dket{1}~
\label{N super R}
\end{eqnarray}
for later convenience. 
It has the following zero-mode part (of bosonic sector)
\begin{eqnarray}
\ket{N}_{R}^{(0)}
= \sum_{w\in \bsz} \ket{\frac{2w+1}{\sqrt{2}}}_L\otimes 
\ket{-\frac{2w+1}{\sqrt{2}}}_R~.
\label{N R 0}
\end{eqnarray}
One might feel our definition of Ramond boundary state peculiar,
since \eqn{N R 0} does not have the standard winding modes
compatible with the compactification $X\sim X+2\sqrt{2}\pi$.
Moreover, $\ket{N}_{NS}+\ket{N}_{R}$ is {\em not\/} the one
describing the BPS $D$-brane. This choice of winding modes,
however, make it possible for the $SU(2)_2$ currents to act locally on
the boundary state  \eqn{N super R} (which is obvious since
we identify $\ket{N}_R$ as the $SU(2)$ Ishibashi state $\dket{1}$),
and it will turn out later that this definition is actually useful 
to describe the S-brane boundary states in superstring.

The Dirichlet boundary states can be written in the similar manner to 
the bosonic case. We again have two possibilities
\begin{eqnarray}
\ket{D}&=& e^{-i\pi J^1_0} \left(\ket{N}_{NS} +\ket{N}_R \right) 
\equiv e^{-i\pi J^1_0} \left(\dket{0}+\dket{2}+\dket{1}\right)~, 
\label{D super SU(2) Ishibashi 1}\\
\ket{D'}&=& e^{i\pi J^1_0} \left(\ket{N}_{NS} +\ket{N}_R \right) 
\equiv e^{i\pi J^1_0} \left(\dket{0}+\dket{2}+\dket{1}\right)  \nonumber\\
&\equiv& e^{-i\pi J^1_0} \left(\dket{0}+\dket{2}-\dket{1}\right)~.
\label{D super SU(2) Ishibashi 2} 
\end{eqnarray}
The former \eqn{D super SU(2) Ishibashi 1} describes 
the alternating $D-\bar{D}$ brane array (under the suitable choice 
of phase factor);
\begin{eqnarray}
\left\{
\begin{array}{ll}
 \mbox{$D$-branes}&~~ \mbox{located at}~ 
X=\frac{2\pi}{\sqrt{2}}\left(2n+\frac{1}{2}\right) ~~(n\in \bz)\\
 \mbox{$\bar{D}$-branes}&~~ \mbox{located at}~ 
X=\frac{2\pi}{\sqrt{2}}\left(2n-\frac{1}{2}\right) ~~(n\in \bz)
\end{array}
\right.
\label{brane array}
\end{eqnarray}
and the latter \eqn{D super SU(2) Ishibashi 2} corresponds 
to the configuration with the inverse brane charges.
We also note that in the open string channel the twisted boundary condition 
\eqn{twisted glue SU(2)} leads to 
\begin{eqnarray}
\psi_L(z)=-\psi_R(\bar{z})~, ~~~
e^{\pm i\sqrt{2}X_L(z)}= -e^{\mp i \sqrt{2}X_R(\bar{z})}~,
\end{eqnarray}
implying the brane array; 
$\dsp X\equiv X_L+X_R=
\frac{2\pi}{\sqrt{2}}\left(n+\frac{1}{2}\right)$ ($n\in \bz$).
Moreover, based  on a careful observation for the RR sector,
we can find the $D-\bar{D}$ configuration \eqn{brane array}
with the expected periodicity $2\sqrt{2}\pi$.


~

\section{Thermal Partition Functions for the S-branes}
\indent

In this section we shall present our main analysis 
of the thermal partition functions for the S-brane backgrounds. 
We only consider the $g_s=0$ limit, in which the desired partition
functions are given as the cylinder amplitudes in the Euclidean signature.
We start with the analysis in bosonic string case.


~

\subsection{Bosonic String Case}
\indent

We first consider  the full S-brane 
background (the bouncing tachyon profile), defined by
the world-sheet action \cite{Sen1};
\begin{eqnarray}
S_L=-\frac{1}{4\pi}\int_{\Sigma} d^2\sigma\, (\partial_{\mu}X^0)^2
 +\la \int_{\partial \Sigma}d\tau\, \cosh X^0~,
\label{S L}
\end{eqnarray}
Based on the periodicity and physical reason discussed in \cite{Sen1}, 
we restrict the range of the coupling $\la$ to $0\leq \la \leq 1/2$
following \cite{LLM}. 
The Wick-rotation $X^0 \, \rightarrow\, iX$ leads to 
the $c=1$ boundary conformal system studied 
in \cite{CKLM,PT,RS} in detail; 
\begin{eqnarray}
S_E= \frac{1}{4\pi}\int_{\Sigma} d^2\sigma\, (\partial_{\mu}X)^2
 +\la \int_{\partial \Sigma}d\tau\, \cos X~.
\label{S E}
\end{eqnarray}
(This is a tautological statement since the solution \eqn{S L}
was presented by Sen by taking the inverse Wick rotation of
\eqn{S E} from the beginning.) As is clarified in \cite{CKLM},
the boundary interaction term is captured by the insertion of 
zero-mode of the $SU(2)$ chiral current; 
\begin{eqnarray}
\la \int_{\partial \Sigma}d\tau\, \cos (X_L+X_R) \sim
\la \oint dz \, \cos(2X_L) = 2\pi i \la J^1_0~, 
\end{eqnarray}
where we used the Neumann boundary condition.
We thus obtain the boundary state for the Euclidean model 
with the compactification $X\sim X+2\pi$ \cite{CKLM}
\begin{eqnarray}
 \ket{B_{\la,1}} \equiv e^{2\pi i \la J^1_0} \ket{N}~,
\end{eqnarray}
where the Neumann boundary state $\ket{N}$ is defined in \eqn{N}.

We can also consider more general compactification 
$X\sim X + 2\pi k$ ($k \in \bz_{>0}$), 
which is compatible with the locality of the $SU(2)$
currents \eqn{SU(2) 1 currents}. This system is interpreted as the 
thermal model with the {\em discretized\/} temperature $T=1/(2\pi k)$
compatible with the S-brane background,  and the relevant 
boundary state  is given \cite{MSY} as
\begin{eqnarray}
\ket{B_{\la,k}} &\equiv &P_k \, e^{2\pi i \la J^1_0}\ket{N} ~, \nonumber \\
   & = & \frac{1}{2^{1/4}}\,
P_k \, e^{2\pi i \la J^1_0} \left(\dket{0}+ \dket{1}\right)~,
\label{B la k}
\end{eqnarray}
where $P_k$ is the projection operator to the Fock space with 
the momenta compatible with the thermal compactification;
\begin{eqnarray}
(p_L,p_R) = \left(\frac{p}{k}+kw, \, \frac{p}{k}-kw \right)~, ~~
(p, w \in \bz )~.
\end{eqnarray}
Note that $k=2$ case corresponds to the Hagedorn temperature
$T_H=1/4\pi$ \cite{Hagedorn}. 
It is obvious that $P_1=\id$ and  the zero-temperature
case $k=\infty$ is described by the projection to the no winding 
Fock space $p_L=p_R$. The corresponding boundary state is the well-known one
\cite{CKLM,PT,RS}
\begin{eqnarray}
\ket{B_{\la,\infty}}&=&P_{\infty }\,
e^{2\pi i \la J^1_0}\left(\dket{0}+\dket{1}\right) \nonumber \\
&=& \sum_{j\in \frac{1}{2}\bsz_{\geq 0}}\sum_{m=-j}^j
 \, D^j_{m,-m}(e^{2\pi i \la \frac{\sigma_1}{2}}) \dket{j,m,m}~,
\label{B la}
\end{eqnarray}
where $D^j_{m,m'}$ denotes the Wigner function (the representation matrix of 
$SU(2)$).
The boundary state of finite temperature
is similarly written as \cite{MSY} 
\begin{eqnarray}
\ket{B_{\la,k}}
&=& \sum_{j\in \frac{1}{2}\bsz_{\geq 0}}\sum_{m,w}
 \, D^j_{m-wk,-m}(e^{2\pi i \la \frac{\sigma_1}{2}}) \dket{j,m,m-wk}~,
\label{B la k 2}
\end{eqnarray}
where the sum of $m$, $w$ should be taken in the range such that
$|m-wk| \leq j$.


~

Now, let us consider our main problem. 
We shall calculate the thermal cylinder amplitude of open strings 
ended at a single s$p$-brane, assuming  the flat space-time 
for simplicity.  
The main part of calculation is the evaluation 
of the overlap along the Euclidean time; 
$\bra{B_{\la,k}}e^{-\pi s H^{(c)}} \ket{B_{\la,k}}$ 
with respect to the boundary states \eqn{B la k} (or \eqn{B la k 2}).
This is in principle calculable, since all the things appearing in 
\eqn{B la k 2} are well-known quantities. 
However, the expressions \eqn{B la k 2} 
are not so useful for the relevant problem,
since the Wigner function $D^j_{m,m'}$ has a quite complicated form 
in general (see for instance \cite{RS}). 
We shall thus take an other route. We calculate the amplitude
based on the Ishibashi states {\em as the 
$SU(2)$ current algebra\/} rather than the Virasoro Ishibashi states. 

To begin with we note the relation 
\begin{eqnarray}
P_k \, e^{2\pi i \la J^1_0} \dket{\ell}&=& 
\frac{1}{k}\sum_{r\in \bsz_k} e^{2\pi i \frac{r}{k}(J^3_0-\tJ^3_0)}
e^{2\pi i \la J^1_0} \dket{\ell} \nonumber \\
&=& \frac{1}{k}\sum_{r\in \bsz_k} e^{2\pi i \frac{r}{k}J^3_0}
e^{2\pi i \la J^1_0} e^{2\pi i \frac{r}{k}J^3_0} \dket{\ell}~,
\label{Pk bosonic}
\end{eqnarray}
which is the key identity for our calculation. 
In the second line we used the boundary condition \eqn{glue SU(2)}.
(We should emphasize that $P_k$ is {\em not\/} equal   
$\dsp \frac{1}{k}\sum_{r\in \bsz_k} 
e^{2\pi i \frac{r}{k}(J^3_0-\tJ^3_0)}$ as an operator.)

Therefore, the overlap can be written as 
\begin{eqnarray}
\bra{B_{\la,k}}e^{-\pi s H^{(c)}} \ket{B_{\la,k}}
&=& \frac{1}{\sqrt{2}}\sum_{\ell =0,1}\dbra{\ell} e^{- 2\pi i \la J^1_0} P_k 
e^{-\pi s H^{(c)}} P_k e^{2\pi i \la J^1_0} \dket{\ell} \nonumber \\
&=& \frac{1}{\sqrt{2}}
\frac{1}{k}\sum_{r\in \bsz_k} \sum_{\ell =0,1} \dbra{\ell} 
e^{-\pi s H^{(c)}}
e^{- 2\pi i \la J^1_0}e^{2\pi i \frac{r}{k}J^3_0}
e^{2\pi i \la J^1_0} e^{2\pi i \frac{r}{k}J^3_0} \dket{\ell}  ~, \nonumber\\
\label{cylinder 1}
\end{eqnarray}
where we used the property of the projection operator; 
$\lb H^{(c)}, \, P_k\rb=0$, $P_k^2=P_k$.

To proceed further we note that
\begin{eqnarray}
&&e^{-2\pi i \la \frac{\sigma_1}{2}} e^{2\pi i z \frac{\sigma_3}{2}}
e^{2\pi i \la \frac{\sigma_1}{2}} e^{2\pi i z \frac{\sigma_3}{2}}
\nonumber \\
&& ~~~= \left(
\begin{array}{cc}
 e^{i\pi z}\left(\cos \left(\frac{\pi r}{k}\right)
+i\sin\left( \frac{\pi r}{k}\right)\cos(2\pi \la)\right)
& 
- e^{-i\pi z} \sin\left(\frac{\pi r}{k}\right) 
\sin\left(2\pi \la\right)
\\
e^{-i\pi z} \sin\left(\frac{\pi r}{k}\right) 
\sin\left(2\pi \la\right)
&
e^{-i\pi z} \left( \cos \left(\frac{\pi r}{k}\right)
-i\sin\left( \frac{\pi r}{k}\right)\cos(2\pi \la)\right)
\end{array}
\right)  \nonumber \\
&& ~~~ = U \left(
\begin{array}{cc}
 e^{2\pi i \frac{1}{2}\al(\la,r/k)} &  0 \\
 0 & e^{-2\pi i \frac{1}{2}\al(\la,r/k)}
\end{array}
\right) U^{-1}~,
\label{calculation 1}
\end{eqnarray}
where  $\al(\la,z)$ is a real number defined by 
\begin{eqnarray}
\al(\la, z) =\frac{2}{\pi} \arcsin \left(\sin(\pi z)\cos(\pi \la)\right)~,
\label{al la}
\end{eqnarray}
and $U$ denotes an unitary matrix (of which explicit form is not
necessary for our argument).
We thus find 
\begin{eqnarray}
e^{- 2\pi i \la J^1_0}e^{2\pi i \frac{r}{k} J^3_0}
e^{2\pi i \la J^1_0} e^{2\pi i \frac{r}{k}J^3_0}
= \cU \, e^{2\pi i \al(\la,r/k)J^3_0} \, \cU^{-1}~,
\label{relation 1}
\end{eqnarray} 
where $\cU$ is an unitary operator of the form such as
$\cU= e^{i\sum_a \theta^a(J^a_0+\tJ^a_0)}
e^{i\sum_a \theta^{a'} (J^a_0+\tJ^a_0)}\cdots ~$.
We also note the identity; $\cU^{-1} \dket{\ell}=\dket{\ell}$
which is obvious from the boundary condition \eqn{glue SU(2)}.

In this way we can evaluate as 
\begin{eqnarray}
\bra{B_{\la,k}} e^{-\pi s H^{(c)}} \ket{B_{\la,k}}
&=& \frac{1}{\sqrt{2}} \frac{1}{k} \sum_{r\in \bsz_{k}} 
 \sum_{\ell =0,1} \dbra{\ell} e^{-\pi s H^{(c)}}
 e^{2\pi i \al(\la,r/k)J^3_0}\dket{\ell}  \nonumber \\
&=& \frac{1}{\sqrt{2}} \frac{1}{k} \sum_{r\in \bsz_{k}} 
 \frac{1}{\eta(is)}\left\{\Theta_{0,1}(is,\al(\la,r/k))
+ \Theta_{1,1}(is,\al(\la,r/k))\right\}  \nonumber \\
&=& \frac{1}{k} \sum_{r\in \bsz_{k}} 
  \frac{1}{\eta(it)} e^{-2\pi t\frac{1}{4}\al(\la,r/k)^2}
 \Theta_{0,1}(it, i\al(\la,r/k)t) \nn
&=&  \frac{1}{k} \sum_{r\in \bsz_{k}} \sum_{n\in \bsz} 
  \frac{1}{\eta(it)} e^{-2\pi t \left(n+\frac{1}{2}\al(\la,r/k)\right)^2}~,
\label{overlap 1} 
\end{eqnarray}
where we have introduced the open string modulus $t\equiv
1/s$ and made use of the familiar modular property of theta function. 

Taking account also of the spatial and ghost sectors, 
we obtain the final result of thermal partition function\footnote
    {As we will clarify later, it turns out that 
     the $t$-integral here includes UV and IR divergences for 
     generic $\la$. Hence we should introduce the UV and IR cut-off's
     into the $t$-integral to be more rigorous.};
\begin{eqnarray}
Z(\la,k)&=& \cN \int_0^{\infty}\frac{dt}{t} \frac{V_p}{(8\pi^2 t)^{p/2}}
\frac{1}{\eta(it)^{24}} \,
\frac{1}{k}\sum_{r\in \bsz_k} 
\sum_{n\in \bsz}
e^{-2\pi t \left(n+\frac{1}{2}\al(\la,r/k)\right)^2}
\label{cylinder amplitude 1} 
\end{eqnarray} 
where $V_p$ is the volume of spatial part of s$p$-brane and 
$\cN$ is the normalization factor which is determined just below. 
Under the zero-temperature limit $k\,\rightarrow\, \infty$, 
the summation of $r$ is replaced with an integral, 
and we obtain 
\begin{eqnarray}
Z(\la,\infty)&=& \cN \int_0^{\infty}\frac{dt}{t} \frac{V_p}{(8\pi^2 t)^{p/2}}
\frac{1}{\eta(it)^{24}} \,
\int_0^1dz\,
\sum_{n\in \bsz}
e^{-2\pi t \left(n+\frac{1}{2}\al(\la,z)\right)^2} ~,
\label{cylinder amplitude zero temperature} 
\end{eqnarray}
which coincides the result already given in \cite{CKLM} (in the 
``note added in proof''  in the NPB version as was pointed out in \cite{MTV}).
See also \cite{PT}.

As a consistency check, let us focus on  the special points 
$\la=0$ and  $\la=1/2$. 
For $\la=0$ we have $\dsp \al(0,r/k)= \frac{2r}{k}$, leading to
\begin{eqnarray}
\bra{B_{0,k}} e^{-\pi s H^{(c)}} \ket{B_{0,k}}
&=& \frac{1}{k} \sum_{r\in \bsz_{k}} 
  \frac{1}{\eta(it)} \sum_{n\in \bsz} e^{-2\pi t
   \left(n+\frac{r}{k}\right)^2} \nn 
&=& \frac{1}{k} \frac{1}{\eta(it)} \sum_{n\in \bsz} 
e^{-2\pi t \frac{n^2}{k^2}} \nn
&=& \frac{1}{k} \frac{1}{\eta(it)} \frac{k}{\sqrt{2t}}
\sum_{n\in \bsz} e^{-\frac{(2\pi k n)^2}{8\pi t}}~,
\end{eqnarray}
where we used the Poisson resummation formula in the last line.
We thus obtain 
\begin{eqnarray}
Z(0,k)= \frac{\cN}{k} \int^{\infty}_0\frac{dt}{t} 
\frac{2\pi k V_p}{(8\pi^2 t)^{\frac{p+1}{2}}} \frac{1}{\eta(it)^{24}}
\sum_{n\in \bsz}e^{-\frac{(2\pi kn)^2}{8\pi t}}~.
\label{cylinder amplitude 2}
\end{eqnarray}
As is expected, this is precisely the 
thermal partition function for a single (time-like)
D$p$-brane with temperature
$T=1/(2\pi k)$ up to normalization. 
Therefore, we should set $\cN=k$ for the correct normalization.

For $\la= 1/2$ we have $\al(1/2, r/k)=0$ and obtain
\begin{eqnarray}
\bra{B_{1/2,k}} e^{-\pi s H^{(c)}} \ket{B_{1/2,k}}
= \frac{\Theta_{0,1}(it,0)}{\eta(it)}~.
\end{eqnarray} 
The thermal partition function now becomes 
\begin{eqnarray}
Z(1/2, k) &=& k \int_0^{\infty}\frac{dt}{t} \frac{V_p}{(8\pi^2 t)^{p/2}}
\frac{1}{\eta(it)^{24}} \sum_{n\in \bsz} e^{-\frac{(2\pi n)^2 t}{2\pi}}~.
\label{cylinder amplitude 3}
\end{eqnarray}
This is of course consistent with the periodic array of  
$D(p-1)$ brane instantons  
at $X=2\pi(n+1/2)$ ($n \in \bz$) mentioned previously. 
Note that they have no support on the real time axis, as is expected
since $\la=1/2$ means that we are already sitting at 
the minimum of tachyon potential \cite{Sen1}.
The overall factor $k$ is consistent with the fact that
we have $k$ $D$-brane instanton on the thermal circle of radius $k$. 
Note that the amplitude in this case does not depend on the 
temperature $T=1/(2\pi k)$ except for the normalization. 
This is due to the simple fact;
$P_k\,\widehat{\dket{\ell}} = \widehat{\dket{\ell}}$ $({}^{\forall} k)$,
and regarded as the first signal of the ``effective thermalization'' 
caused by S-brane discussed in \cite{Strominger,GS2,MSY}.

~


The calculation for the half S-brane is almost parallel.
We shall only focus on the brane decay solution 
$\dsp T(X^0)=\la e^{X^0}$. 
The world-sheet action is given by
\begin{eqnarray}
S_L=-\frac{1}{4\pi}\int_{\Sigma} d^2\sigma\, (\partial_{\mu}X^0)^2
 +\la \int_{\partial \Sigma}d\tau\, e^{X^0}~.
\label{S L 2}
\end{eqnarray}
The thermal boundary state is similarly constructed 
by inserting $e^{2\pi i \la J^+_0}$;
\begin{eqnarray}
\ket{B^{(+)}_{\la,k}}=\frac{1}{2^{1/4}}\,
P_k\,e^{2\pi i \la J^+_0} \left(\dket{0}+\dket{1}\right)~.
\label{B la k half} 
\end{eqnarray}
The calculation of cylinder amplitude can be carried out
in the same way. The most non-trivial point is 
\begin{eqnarray}
&&e^{- 2\pi i \la J^-_0}e^{2\pi i z J^3_0}
e^{2\pi i \la J^+_0} e^{2\pi i z J^3_0}
\sim e^{2\pi i \al^{(+)}(\la,r)J^3_0} ~,  \nn
&& 
\al^{(+)}(\la,z) = \frac{1}{\pi}\arccos\left(\cos(2\pi z) +
\frac{1}{2} (2\pi\la)^2\right)~,
\label{al + la}
\end{eqnarray} 
where $\sim $ means the equality up to a similarity transformation 
as in \eqn{relation 1}.
Notice that $\al^{(+)}(\la, z)$ here is generally 
a complex number contrary to the full brane case. 
This is because $e^{2\pi i \la J^+_0}$ is not an unitary operator. 
To be more specific, we find 
\begin{eqnarray}
\al^{(+)}(\la,z) \in \br && ~~ \mbox{if} ~~ 
\left|\cos(2\pi z)+\frac{1}{2}(2\pi \la)^2\right| \leq 1 ~, \nn
\al^{(+)}(\la,z) \in i \br && ~~ \mbox{if} ~~ 
\left|\cos(2\pi z)+\frac{1}{2}(2\pi \la)^2\right| > 1 ~.
\label{evaluation al +} 
\end{eqnarray}
The final result is written as 
\begin{eqnarray}
Z_{\msc{half}}(\la,k)
&=& k \int_0^{\infty}\frac{dt}{t} \frac{V_p}{(8\pi^2 t)^{p/2}}
\frac{1}{\eta(it)^{24}} \,
\frac{1}{k}\sum_{r\in \bsz_k} 
\sum_{n\in \bsz} e^{-2\pi t \left(n+\frac{1}{2}\al^{(+)}(\la,r/k)\right)^2}
~,
\label{cylinder amplitude half} 
\end{eqnarray}
This is a real function as we can check it by using \eqn{evaluation al +}.

We here make an important comment. 
In the world-sheet action \eqn{S L 2}
we can absorb the coupling $\la$ into the zero-mode of $X^0$.
In other words, the coupling $\la$ is {\em dynamical\/} in  the
Lorentzian theory and should be integrated out when performing
the path integration. In fact, the Lorentzian cylinder amplitudes 
calculated in \cite{LLM,KLMS} do not depend on $\la$ as should be.  
On the other hand, the path integration is 
performed along the imaginary time axis in our thermal model. 
Rewriting as   $\la\equiv \la_0 e^{x^0}$, 
we can regard the {\em real\/} time coordinate $x^0$ as a {\em parameter\/}
of Euclidean world-sheet action. 
The thermal partition function \eqn{cylinder amplitude half}
now gains  an explicit time dependence through the coupling $\la$.


~

\subsection{Superstring Case}
\indent

Next let us turn to the superstring case.
First we again consider the full S-brane, which is defined as
the non-BPS D-brane with the tachyon profile 
$\dsp T(X^0) = \frac{\la}{\sqrt{2}} \cosh (X^0/\sqrt{2})$.
The corresponding vertex operator in the zero-picture 
is given \cite{Sen2} as 
\begin{eqnarray}
\frac{\la}{\sqrt{2}}
 \int_{\partial \Sigma} d\tau\, \sigma_1\otimes \psi^0 \sinh (X^0/\sqrt{2})~,
\end{eqnarray}
where $\sigma_1$ is the Chan-Paton (CP) factor characteristic for the non-BPS
D-brane. The Wick-rotated world-sheet action is given by
\begin{eqnarray}
S_E&=& \frac{1}{4\pi}\int_{\Sigma} d^2\sigma\, (\partial_{\mu}X)^2
 -\frac{\la}{\sqrt{2}} 
\int_{\partial \Sigma}d\tau\, \sigma_1\otimes \psi \sin(X/\sqrt{2})~, \nn
&=& \frac{1}{4\pi}\int_{\Sigma} d^2\sigma\, (\partial_{\mu}X)^2
 +i\frac{\la}{2\sqrt{2}} 
\int_{\partial \Sigma}d\tau\, \sigma_1\otimes \psi 
( e^{iX/\sqrt{2}}- e^{-iX/\sqrt{2}})~.
\label{S E 2}
\end{eqnarray}
Hence the boundary interaction is again described by the $SU(2)_2$
current \eqn{SU(2) 2 currents}. However, the problem is slightly non-trivial
compared with the bosonic case because of the existence of 
CP matrix $\sigma_1$. The boundary interaction should be realized as
an insertion of Wilson line \cite{CLNY}.  
For the NSNS sector, the Wilson line is simply given as
\begin{eqnarray}
\frac{1}{2} \tr \, \left(e^{2\pi i \la \sigma_1\otimes J^1_0}\right) = 
\frac{1}{2}
\left(e^{2\pi i \la J^1_0} +e^{-2\pi i \la J^1_0}\right)~. 
\label{W L NS}
\end{eqnarray}
For the RR sector, on the other hand, we need to be more careful.
We must insert the extra CP factor  $\sigma_1$ into the trace so as 
to obtain the consistent result (see \cite{Sen-review}, for instance);
\begin{eqnarray}
\frac{1}{2}\tr \, \left(\sigma_1\,e^{2\pi i \la \sigma_1\otimes J^1_0}\right) 
=\frac{1}{2} 
\left(e^{2\pi i \la J^1_0} -e^{-2\pi i \la J^1_0}\right)~. 
\label{W L R}
\end{eqnarray}

As in the bosonic case, we can consider the thermal compactification
$X\sim X+2\pi\sqrt{2}k$ ($k\in \bz_{>0}$) compatible with 
the boundary interaction. The zero-mode momenta should be restricted as
\begin{eqnarray}
(p_L,p_R)=\left(\frac{p}{\sqrt{2}k}+\sqrt{2}kw,\, 
\frac{p}{\sqrt{2}k}-\sqrt{2}kw\right)~, ~~ (p,w\in\bz)~,
\end{eqnarray}
and we denote the associated thermal projection operator as $P_k$.
Taking account of the Wilson line operators, 
the thermal boundary state is written as
\begin{eqnarray}
\ket{B_{\la,k}} &=& 
\frac{1}{2} P_k\, (e^{2\pi i \la J^1_0} +e^{-2\pi i \la J^1_0}) 
\sqrt{2}\ket{N}_{NS}
+ \frac{1}{2}P_k\,(e^{2\pi i \la J^1_0} -e^{-2\pi i \la J^1_0}) 
\sqrt{2}\ket{N}_{R} \nn
&\equiv& \frac{1}{\sqrt{2}}P_k\,(e^{2\pi i \la J^1_0} +e^{-2\pi i \la J^1_0})
\left(\dket{0}+\dket{2}\right)
+ \frac{1}{\sqrt{2}}
P_k\,(e^{2\pi i \la J^1_0} -e^{-2\pi i \la J^1_0}) \dket{1}~,
\label{B la k super}
\end{eqnarray}
where the factor $\sqrt{2}$ is originating from  
the tension of the non-BPS D-brane. 
To confirm the validity of it 
we first point out that, under  the zero temperature limit 
$k\rightarrow \infty$,  the boundary state \eqn{B la k super}
correctly reproduces the NSNS and RR source terms of zero-mode
sectors (in the Lorentzian theory) presented in
\cite{Sen2,Sen3};
\begin{eqnarray}
&& f_{\sNS}(x^0)= \frac{1}{1+e^{\sqrt{2}x^0}\sin^2(\pi\la)}
+ \frac{1}{1+e^{-\sqrt{2}x^0}\sin^2(\pi\la)} -1~, \nn
&& f_{\sR}(x^0) = \sin(\pi \la) \left(
\frac{e^{\frac{x^0}{\sqrt{2}}}}{1+e^{\sqrt{2}x^0}\sin^2(\pi \la)}
-\frac{e^{-\frac{x^0}{\sqrt{2}}}}{1+e^{-\sqrt{2}x^0}\sin^2(\pi \la)}
\right)~.
\label{source}
\end{eqnarray}
Secondly, let us  focus on 
the special points $\la=0$ and $\la=1/2$ (or $\la=-1/2$).
It is easily found that
\begin{eqnarray}
\ket{B_{0,k}}&=& \sqrt{2} P_k\, \ket{N}_{NS}
\equiv \sqrt{2} P_k\,( \dket{0}+\dket{2})~, \nn
\ket{B_{1/2,k}}&=& \sqrt{2} e^{-i\pi J^1_0} (\ket{N}_{NS}+\ket{N}_{R})
  \equiv \sqrt{2} e^{-i\pi J^1_0}( \dket{0} + \dket{2}+\dket{1}) 
\equiv \sqrt{2} \ket{D} ~, \nn
\ket{B_{-1/2,k}}&=& \sqrt{2} e^{-i\pi J^1_0} (\ket{N}_{NS}-\ket{N}_{R})
\equiv \sqrt{2} 
e^{-i\pi J^1_0}( \dket{0} + \dket{2}-\dket{1}) 
 \equiv \sqrt{2} \ket{D'}~.
\label{interpolation}
\end{eqnarray}
(Recall the relations  \eqn{D super SU(2) Ishibashi 1}, 
\eqn{D super SU(2) Ishibashi 2}.)
The $\la =0 $ case correctly reproduces the thermal boundary state of 
the non-BPS $D$-brane, and the $\la =\pm 1/2$ cases correspond  to
the periodic $D-\bar{D}$ instanton configurations as is expected. 
In fact, if neglecting the projection $P_k$, 
the boundary state \eqn{B la k super} is precisely 
the T-dual of the one describing the interpolation between 
the non-BPS $D0$-brane and $D1-\bar{D}1$ system 
(with the $\bz_2$ Wilson line) compactified on the critical radius 
$R=\sqrt{2}$ \cite{Sen-old1,FGLS,Sen-review}. (This is again a 
tautological statement by construction of the rolling tachyon solution.)
However,  the existence of $P_k$ is essential for our analysis and 
makes the cylinder amplitude quite different from that given in 
\cite{FGLS}.  
Note also that we have no $k$ dependence when $\la=\pm 1/2$.
This is again originating from the fact that the thermal projection
$P_k$ trivially acts on the Dirichlet boundary states.

~

Now we are in the position to calculate the thermal partition function.
To make use of the similar technique as in the bosonic case,
we first note that   
\begin{eqnarray}
P_k \, e^{2\pi i \la J^1_0} \dket{\ell}&=& 
\frac{1}{2k}\sum_{r\in \bsz_{2k}} e^{2\pi i \frac{r}{2k}(J^3_0-\tJ^3_0)}
e^{2\pi i \la J^1_0} \dket{\ell} \nonumber \\
&=& \frac{1}{2k}\sum_{r\in \bsz_{2k}} e^{2\pi i \frac{r}{2k}J^3_0}
e^{2\pi i \la J^1_0} e^{2\pi i \frac{r}{2k}J^3_0} \dket{\ell}~,
\label{Pk super}
\end{eqnarray}
which is similar to \eqn{Pk bosonic} but includes  slightly different
coefficients. 
In addition to \eqn{relation 1}, \eqn{al la}, the following relation 
is useful;
\begin{eqnarray}
\left(e^{2\pi i \la J^1_0}e^{2\pi i zJ^3_0}\right)^2
\sim e^{2\pi i \beta(\la,z)J^3_0}~, 
\label{relation 2}
\end{eqnarray} 
where we set
\begin{eqnarray}
\beta(\la,z)=\frac{2}{\pi} \arccos\left(\cos(\pi z)\cos(\pi \la)\right)
~.
\label{beta la}
\end{eqnarray}
Another non-trivial point is the thermal boundary condition for the
fermionic sector given in \cite{AW}. In our present problem
it amounts to the insertion of the ``thermal GSO projection''
\begin{eqnarray}
 \frac{1}{2}\left(1+(-1)^{F_L+W}\right)\equiv  
\frac{1}{2}\left(1+(-1)^{F_L}e^{2\pi i \frac{1}{4k}(J^3_0-\tJ^3_0)}\right)~,
\label{thermal GSO}
\end{eqnarray}
where $F_L$ denotes the fermion number operator in the left mover and 
$W\equiv p_L-p_R$ stands for the winding along the thermal circle,  
instead of the usual GSO projection 
$\dsp \frac{1}{2}\left(1+(-1)^{F_L}\right)$ \footnote
   {When calculating the overlap of boundary states, it is enough to 
    consider the GSO projection only for the left mover
     (or the right mover) $\dsp \frac{1}{2}\left(1+(-1)^{F_L}\right)$
     in place of  $\dsp \frac{1}{2}\left(1+(-1)^{F_L}\right)\cdot
     \frac{1}{2}\left(1+(-1)^{F_R}\right)$. }.
The following formulas are also useful;
\begin{eqnarray}
&& \lb (-1)^{F_L}e^{i\pi J^3_0},\, J^a_n \rb =0~,~~~ ({}^{\forall}a, n) ~,\nn
&& (-1)^{F_L}e^{i\pi J^3_0} \dket{0}= -\dket{0}~, ~~~
   (-1)^{F_L}e^{i\pi J^3_0} \dket{2}= \dket{2}~, \nn
&& (-1)^{F_L}\dket{1}=0~.
\label{formulas 1}
\end{eqnarray}
(We take the standard convention
$(-1)^{F_L}\ket{0}_{\sNS}=-\ket{0}_{\sNS}$
for the NS Fock vacuum.) The last identity is due to the fermion
zero-modes in the Ramond sector.

In this way we can evaluate the overlap amplitude as follows;
\begin{eqnarray}
&& {}_{\sNS}\bra{B_{\la,k}} 
e^{-\pi s H^{(c)}} \ket{B_{\la,k}}_{\sNS} \nn
&& ~~~= \frac{1}{2k}\sum_{r\in \bsz_{2k}}\sum_{\ell =0,2}
\left\{\dbra{\ell}e^{-\pi s H^{(c)}}  e^{2\pi i \al(\la,\frac{r}{2k})J^3_0}
\dket{\ell}  + 
\dbra{\ell}e^{-\pi s H^{(c)}}  e^{2\pi i \beta(\la,\frac{r}{2k})J^3_0}
\dket{\ell}  \right. \nn
&& ~~~\hspace{1cm} + \left. \dbra{\ell}e^{-\pi s H^{(c)}}  
e^{2\pi i \al(-\la,\frac{r}{2k})J^3_0}
\dket{\ell}  + 
\dbra{\ell}e^{-\pi s H^{(c)}}  e^{2\pi i \beta(-\la,\frac{r}{2k})J^3_0}
\dket{\ell}  \right\} \nn
&& ~~~= \frac{1}{2k}
\sum_{r\in \bsz_{2k}}\sum_{\ell =0,2} \left\{
\chi^{(2)}_{\ell}(is,\al(\la, \frac{r}{2k})) +
\chi^{(2)}_{\ell}(is,\beta(\la, \frac{r}{2k})) 
\right\} \nn
&& ~~~ = \frac{1}{2k}
\sum_{r\in \bsz_{2k}}\left\{
\sqrt{\frac{\th_3(is,0)}{\eta(is)}} 
\frac{\th_3(is,\al(\la, \frac{r}{2k}))}{\eta(is)}
+\sqrt{\frac{\th_3(is,0)}{\eta(is)}} 
\frac{\th_3(is,\beta(\la, \frac{r}{2k}))}{\eta(is)}
\right\}~.
\label{overlap NS}
\end{eqnarray}
We likewise obtain for the RR sector
\begin{eqnarray}
&& {}_{\sR}\bra{B_{\la,k}} e^{-\pi s H^{(c)}} \ket{B_{\la,k}}_{\sR} \nn
&& ~~~= -\frac{1}{2k} \sum_{r\in\bsz_{2k}}
\left\{   \sqrt{\frac{\th_2(is,0)}{2\eta(is)}} 
\frac{\th_2(is,\al(\la,\frac{r}{2k}))}{\eta(is)} -
\sqrt{\frac{\th_2(is,0)}{2\eta(is)}} 
\frac{\th_2(is,\beta(\la,\frac{r}{2k}))}{\eta(is)}
  \right\}~.
\label{overlap R}
\end{eqnarray}
The calculation of ${}_{\sNS}\bra{B_{\la,k}} 
  (-1)^{F_L}e^{2\pi i \frac{1}{4k}(J^3_0-\tJ^3_0)} 
e^{-\pi s H^{(c)}} \ket{B_{\la,k}}_{\sNS}$ is more non-trivial.
With the helps of \eqn{formulas 1} we can  
make use of the identities  such as
\begin{eqnarray}
&& \dbra{0}e^{-2\pi i \la J^1_0}(-1)^{F_L}
= \dbra{0}(-1)^{F_L}e^{2\pi i \la J^1_0} 
= -\dbra{0} e^{i\pi J^3_0}e^{2\pi i \la J^1_0}~, \nn
&& \dbra{2}e^{-2\pi i \la J^1_0}(-1)^{F_L}
= \dbra{2}(-1)^{F_L}e^{2\pi i \la J^1_0} 
= \dbra{2} e^{i\pi J^3_0}e^{2\pi i \la J^1_0}~
\end{eqnarray}
and also introduce
\begin{eqnarray}
&& e^{i\pi J^3_0} e^{-2\pi i \la J^1_0} e^{2\pi i z J^3_0 }
e^{2\pi i \la J^1_0} e^{2\pi i z J^3_0 } 
\sim e^{i\pi J^3_0} (e^{2\pi i \la J^1_0} e^{2\pi i z J^3_0 })^2 
\sim  e^{2\pi i \gamma(\la, z) J^3_0}~, \nn
&&
\gamma(\la,z) =\frac{1}{\pi} \arccos 
\left(\cos\left(\pi\left(2z+\frac{1}{2}\right)\right)\cos^2(\pi \la)\right)~, 
\end{eqnarray}
as in \eqn{relation 1}, \eqn{relation 2}.
We thus obtain 
\begin{eqnarray}
&&{}_{\sNS}\bra{B_{\la,k}} 
  (-1)^{F_L}e^{2\pi i \frac{1}{4k}(J^3_0-\tJ^3_0)} 
e^{-\pi s H^{(c)}} \ket{B_{\la,k}}_{\sNS}  \nn
&& ~~~ = \frac{1}{2k}\sum_{r\in \bsz_{2k}}\left\{
-\chi^{(2)}_0(is,\gamma(\la, \frac{r+1/2}{2k}))+
\chi^{(2)}_2(is,\gamma(\la, \frac{r+1/2}{2k}))
\right\}  \nn
&& ~~~ = - \frac{1}{2k}\sum_{r\in \bsz_{2k}}
\sqrt{\frac{\th_4(is,0)}{\eta(is)}}
\frac{\th_4(is,\gamma(\la, \frac{r+1/2}{2k}))}{\eta(is)}~.
\label{overlap NS 2}
\end{eqnarray}

Gathering contributions from all the sectors 
and performing the modular transformation
to the open string modulus
$t\equiv 1/s$, we finally obtain the following thermal partition function;
\begin{eqnarray}
&& Z(\la,k)=\cN \int_0^{\infty} \frac{dt}{t}\frac{V_p}{(8\pi^2 t)^{p/2}} 
\frac{1}{\eta(it)^8} \frac{1}{2k}\sum_{r\in \bsz_{2k}} \nn
&& ~~~ \times \frac{1}{2}\left\lb \left(\frac{\th_3}{\eta}\right)^4(it)
\cdot \frac{1}{2}\left\{\th_3(it,i\al(\la,\frac{r}{2k})t)
e^{-2\pi t \frac{1}{2}\al(\la,\frac{r}{2k})^2} +
\th_3(it,i\beta(\la,\frac{r}{2k})t)
e^{-2\pi t \frac{1}{2}\beta(\la,\frac{r}{2k})^2}
\right\} \right. \nn
&& ~~~  -\left(\frac{\th_4}{\eta}\right)^4(it)
\cdot \frac{1}{2}\left\{\th_4(it,i\al(\la,\frac{r}{2k})t)
e^{-2\pi t \frac{1}{2}\al(\la,\frac{r}{2k})^2} -
\th_4(it,i\beta(\la,\frac{r}{2k})t)
e^{-2\pi t \frac{1}{2}\beta(\la,\frac{r}{2k})^2}  
\right\} \nn
&& ~~~ \left. - \left(\frac{\th_2}{\eta}\right)^4(it)
\th_2(it,i\gamma(\la,\frac{r+1/2}{2k})t) e^{-2\pi t \frac{1}{2}
\gamma(\la, \frac{r+1/2}{2k})^2} \right\rb ~,
\label{cylinder amplitude super 1}
\end{eqnarray}
where $\cN$ is a normalization constant which should be determined 
by the consistency with the $\la=0$ case as in the  bosonic case.

Let us consider the special cases $\la=0$, $\la=1/2$ which simplifies
the amplitude. 
For $\la=0$, since $\al(0,z)=2z$, $\beta(0,z)=2z$, 
$\gamma(0,z)=2z+1/2$ holds, we can easily find
\begin{eqnarray}
&& Z(0,k) = \frac{\cN}{2} \int_0^{\infty}  \frac{dt}{t}
\frac{V_p}{(8\pi^2t)^{p/2}}
\frac{1}{\eta(it)^8} \frac{1}{2k}\sum_{r\in \bsz_{2k}} \nn
&& ~~~ \times \left\lb \left(\frac{\th_3}{\eta}\right)^4(it) 
\th_3(it, i\frac{r}{k}t)e^{-2\pi t \frac{1}{2}\left(\frac{r}{k}\right)^2}
-\left(\frac{\th_2}{\eta}\right)^4(it)
\th_3(it, i\frac{r+1/2}{k}t)e^{-2\pi t \frac{1}{2}\left(\frac{r}{k}\right)^2}
\right\rb \nn
&& ~~~= \frac{\cN}{2k} \int_0^{\infty} \frac{dt}{t}\frac{2\sqrt{2}\pi k V_p}
{(8\pi^2t)^{\frac{p+1}{2}}} \frac{1}{\eta(it)^8} 
\sum_{n\in \bsz}e^{-\frac{(2\sqrt{2}\pi k n)^2}{8\pi t}}
\left\{\left(\frac{\th_3}{\eta}\right)^4(it)-
(-1)^n\left(\frac{\th_2}{\eta}\right)^4(it)\right\}~.
\label{cylinder amplitude super 2}
\end{eqnarray}
In the second line we have used the Poisson resummation formula.
This is the correct form of thermal partition function for 
the non-BPS brane with the temperature $T=1/(2\sqrt{2}\pi k)$
up to normalization. Especially, the correct phase factor $(-1)^n$
for the space-time fermions is successfully realized. 
In this way we should fix the normalization constant as $\cN=2k$.

For $\la=1/2$, we have $\al(1/2, z)=0$, $\beta(1/2,z)=1$, and 
$\gamma(1/2, z)=1/2$. We thus obtain after  a little  calculation
\begin{eqnarray}
&& Z(1/2,k)= 2k \int^{\infty}_0 \frac{dt}{t} 
\frac{V_p}{(8\pi^2 t)^{p/2}} \frac{1}{\eta(it)^8} 
\left(\frac{\th_4}{\eta}\right)^4(it) \sum_{n\in \bsz} 
e^{-\frac{t}{2\pi}\left\{2\sqrt{2}\pi (n+1/2)\right\}^2}~.
\label{cylinder amplitude super 3}
\end{eqnarray}
We here used the familiar Jacobi identity
\begin{eqnarray}
\left(\frac{\th_3}{\eta}\right)^4-\left(\frac{\th_4}{\eta}\right)^4
-\left(\frac{\th_2}{\eta}\right)^4=0~.
\end{eqnarray}
It is a straightforward task to confirm the amplitude 
\eqn{cylinder amplitude super 3} is really the one expected from 
the $D(p-1)-\bar{D}(p-1)$ system \eqn{brane array}. The $D-D$ and 
$\bar{D}-\bar{D}$ open strings do not contribute to the amplitude
by the SUSY cancellation, and the $D-\bar{D}$ strings
yield the correct instanton action 
$\dsp \frac{t}{2\pi}\left\{2\sqrt{2}\pi (n+1/2)\right\}^2$.
It may be amazing that the alternating structure of $D-\bar{D}$ system
correctly reproduces the thermal boundary condition for fermions 
(that is implicit in the $\th_4$ factors in \eqn{cylinder amplitude super 3}).
The overall factor $2k$ is the correct degeneracy of configuration
on the thermal circle with radius $\sqrt{2}k$.

~

The thermal partition functions for the half S-branes 
are similarly computed  with a little modification.
We again only consider the 
brane decay case $T(X^0) = \frac{1}{2\sqrt{2}}
\la e^{X^0/\sqrt{2}}$ (in the $(-1)$ picture).
In addition to the ``twist angle''  $\al^{(+)}(\la, z)$ defined in 
\eqn{al + la}, we introduce $\beta^{(+)}(\la,z)$, $\gamma^{(+)}(\la,z)$
as follows; 
\begin{eqnarray}
&&e^{2\pi i \la J^-_0}e^{2\pi i z J^3_0} 
e^{2\pi i \la J^+_0} e^{2\pi i z J^3_0}
\sim e^{2\pi i \beta^{(+)}(\la,z)J^3_0} ~, \nn
&& \beta^{(+)}(\la,z) = \frac{1}{\pi}\arccos\left(\cos(2\pi z)-
\frac{1}{2}(2\pi \la)^2\right)~,
%
\label{beta + la} \\
&& e^{i\pi J^3_0}e^{2\pi i \la J^-_0}e^{2\pi i z J^3_0}
 e^{2\pi i \la J^+_0} e^{2\pi i z J^3_0}
\sim e^{2\pi i \gamma^{(+)}(\la,z) J^3_0}~,\nn
&& \gamma^{(+)}(\la,z) = \frac{1}{\pi}\arccos
\left(\cos\left(\pi\left(2z+\frac{1}{2}\right) \right)+
\frac{i }{2}(2\pi \la)^2\right)~,
\label{gamma + la} \\
&& \left( e^{i\pi J^3_0}e^{-2\pi i \la J^-_0}e^{2\pi i z J^3_0}
 e^{2\pi i \la J^+_0} e^{2\pi i z J^3_0}
\sim e^{2\pi i \gamma^{(+)*}(\la,z)J^3_0} \right) ~.
\nonumber
\end{eqnarray}
The final result is written as
\begin{eqnarray}
&& Z_{\msc{half}}(\la,k)=
2k \int_0^{\infty}  \frac{dt}{t} \frac{V_p}{(8\pi^2 t)^{p/2}} 
\frac{1}{\eta(it)^8} \frac{1}{2k}\sum_{r\in \bsz_{2k}} \nn
&&  \times \frac{1}{2}\left\lb \left(\frac{\th_3}{\eta}\right)^4(it)
\cdot \frac{1}{2}\left\{\th_3(it,i\al^{(+)}(\la,\frac{r}{2k})t)
e^{-2\pi t \frac{1}{2}\al^{(+)}(\la,\frac{r}{2k})^2} +
\th_3(it,i\beta^{(+)}(\la,\frac{r}{2k})t)
e^{-2\pi t \frac{1}{2}\beta^{(+)}(\la,\frac{r}{2k})^2}
\right\} \right. \nn
&& ~~~  -\left(\frac{\th_4}{\eta}\right)^4(it)
\cdot \frac{1}{2}\left\{\th_4(it,i\al^{(+)}(\la,\frac{r}{2k})t)
e^{-2\pi t \frac{1}{2}\al^{(+)}(\la,\frac{r}{2k})^2} -
\th_4(it,i\beta^{(+)}(\la,\frac{r}{2k})t)
e^{-2\pi t \frac{1}{2}\beta^{(+)}(\la,\frac{r}{2k})^2}  
\right\} \nn
&& ~~~ \left. - \left(\frac{\th_2}{\eta}\right)^4(it) \cdot
\frac{1}{2} \left\{
\th_2(it,i\gamma^{(+)}(\la,\frac{r+1/2}{2k})t) e^{-2\pi t \frac{1}{2}
\gamma^{(+)}(\la, \frac{r+1/2}{2k})^2} \right.  \right. \nn
&& \hspace{6cm} \left. \left.
+\th_2(it,i\gamma^{(+)*}(\la,\frac{r+1/2}{2k})t) e^{-2\pi t\frac{1}{2}
\gamma^{(+)*}(\la, \frac{r+1/2}{2k})^2} 
\right\} \right\rb ~. 
\label{cylinder amplitude super half}
\end{eqnarray}
Despite the complexity of  
the angles $\al^{(+)}(\la,z)$, $\beta^{(+)}(\la,z)$ and 
$\gamma^{(+)}(\la,z)$, this is a real function as we can readily confirm it.

~

\section{Effective Hagedorn Behaviors in the S-brane Backgrounds}
\indent

In this section we argue on the UV behaviors of thermal partition
functions we calculated. We first study the full S-brane 
in a detail, and later discuss the half S-brane case.


~

\subsection{Full S-brane in Bosonic String Case}
\indent

We start with focusing on the special points $\la=0$ and $\la=1/2$.
The former is nothing but the ordinary time-like $Dp$-brane.
The thermal amplitude \eqn{cylinder amplitude 2} is identified 
as the free energy of the open string gas attached at the $Dp$-brane
by the simple relation \cite{Polchinski};
\begin{eqnarray}
F(\beta = 2\pi k)&\equiv &\frac{1}{\beta}
\tr \left\lb (-1)^{\msc{\bf F}}
\ln \left(1-(-1)^{\msc{\bf F}}e^{-\beta p^0}\right)\right\rb
= -\frac{1}{\beta}Z(\la=0, k)
\label{free energy}
\end{eqnarray}
Here ${\bf F}$ denotes the space-time fermion number.
$p^0$ is the space-time energy related to 
the open string Hamiltonian $H^{(o)}$ by the on-shell
condition as follows; 
\begin{eqnarray}
p^0 \equiv \frac{1}{\sqrt{2}}(p^+-p^-) 
= \frac{p^+}{\sqrt{2}}+ \frac{H^{(o)}}{2\sqrt{2}p^+}~.
\end{eqnarray}
With the help of this identity
the second equality in \eqn{free energy} can be directly checked 
by taking the light-cone gauge. (The light-cone momentum $p^+$ is converted 
to the cylinder modulus by the suitable change of variable.)
This amplitude \eqn{free energy} includes contributions from all the on-shell 
open string states. 
On the other hand, in the closed string picture, 
the boundary state only includes the off-shell states\footnote
    {The boundary state is of course constructed as to be BRST
   invariant in the corresponding Lorentzian theory; 
   $(Q_{\msc{BRST}}+\tilde{Q}_{\msc{BRST}})\ket{B}=0$ 
   (or, equivalently, to be conformally  invariant in the sense of 
   $(L_n-\tL_{-n})\ket{B}=0$). The ``off-shell'' here 
   means that all the zero-mode momenta appearing 
   in $\ket{B}$ cannot satisfy the mass shell condition 
   for the closed string modes (except for the closed string tachyon). 
   This fact is obvious from the kinematical reason.}.
It is nevertheless quite useful for the investigation of thermodynamical 
behavior of {\em physical\/} open string modes. 
The Hagedorn temperature \cite{Hagedorn} is characterized as
the temperature over which the {\em unphysical\/} closed strings 
wrapped around the thermal circle become tachyonic \cite{Sath,AW}.

For the latter case, $\la=1/2$, the situation is in a sense reversed.  
The background reduces to the space-like $D(p-1)$-brane array  
(``$sD$-brane'' in the terminology of \cite{MSY}) along 
the thermal circle (the imaginary time axis) with interval $2\pi$. 
The cylinder amplitude \eqn{cylinder amplitude 3}
includes on-shell closed string states, but only off-shell states
in the open string channel.
This feature is likely to be consistent with the fact that  
we have no physical $D$-brane on the real time axis at this
point.\footnote
   {The interpretation of $D$-brane instantons located along the
   imaginary time axis as closed string vertices  has been
   discussed in a detail in the paper \cite{GIR}.}
The thermal behavior of physical closed string 
excitations is captured  by the ``virtual'' open strings 
wrapped around the thermal circle. 

Based on these observations we shall concentrate on (i) near the point 
$\la=0$  and (ii) near $\la=1/2$.
We should argue on the thermal behavior of 
open string modes in the first case, and the closed string modes  
in the second case. 

~

\noindent
{\bf (i) Near $\la=0$ : Effective Hagedorn Behavior of 
Open String Excitations}

As we mentioned above, the simplest way to investigate 
the thermal behavior of physical open string excitations 
is to observe the IR behavior of the ``dual'' closed string 
picture. By considering the $q$-expansion of theta functions, 
we find that the thermal partition function \eqn{cylinder amplitude 1}
has the following structure in the closed string channel
(see the second line in \eqn{overlap 1});  
\begin{eqnarray}
Z(\la,k)&=& \int_0^{\infty} ds \, Z^{(c)}(\la,k;s)  ~,\nn
Z^{(c)}(\la,k;s) &=& \mbox{const.}\times \frac{1}{s^{(25-p)/2}} 
\sum_{w=0}^{\infty} Z_w^{(c)}(\la,k;s) ~,
\label{Z c}
\end{eqnarray}
where $w$ denotes the (absolute value of) ``winding number''
along the thermal circle originating from the theta functions 
$\Theta_{*,1}(is,\al(\la,r/k))$
in \eqn{overlap 1} and $Z_w^{(c)}(\la,k;s)$ has the form as
\begin{eqnarray}
Z_w^{(c)}(\la,k;s) = e^{2\pi s \left(1- \frac{w^2}{4}\right)}
\times \mbox{power series of $e^{-2\pi s}$}~.
\label{Z w c}
\end{eqnarray}
It is obvious that the leading term $Z_0^{(c)}(\la,k;s)$ does not depend on 
the parameters $\la$, $k$ and behaves as 
\begin{eqnarray}
Z_0^{(c)}(\la,k;s) = e^{2\pi s} + \cO(1)~, ~~  (s\, \rightarrow \, \infty)
\end{eqnarray}
which corresponds to the closed string tachyon and is expected to
be absent in the superstring case. 
We are interested in $Z_w^{(c)}(\la,k;s)$ ($w\neq 0$).
The first non-zero contribution from them captures  the desired 
thermal behavior of open string excitations.

Let us first set $\la=0$. Since $\dsp \frac{1}{2}\al(0,r/k)=r/k$ holds, 
the ``averaging'' of the form 
$\dsp \frac{1}{k}\sum_{r\in \bsz} f(r/k)$ kills all the sectors of
$w<k$; i.e.
$Z_1^{(c)}(0,k;s)= \cdots =Z_{k-1}^{(c)}(0,k;s)=0$,
and the first non-zero term is $Z_k^{(c)}(0,k;s)$.
This fact reproduces the thermal compactification 
$X\sim X+2\pi k$. To be more precise, 
we obtain  the correct IR behavior at temperature $T=1/(2\pi k)$; 
\begin{eqnarray}
Z^{(c)}(0,k;s) - \frac{1}{s^{(25-p)/2}}
Z^{(c)}_0(0,k;s)\sim \frac{1}{s^{(25-p)/2}} 
e^{2\pi s\left(1-\frac{k^2}{4}\right)} ~, ~~ (s\, \rightarrow \, \infty)
\end{eqnarray}  
and no thermal instability appears, if $k > 2$.
(Recall that $k=2$ corresponds to the Hagedorn temperature.)

Turning on the small coupling $\la$, the situation is drastically changed.
In fact, we can evaluate $Z^{(c)}_1(\la,k;s)$ as 
\begin{eqnarray}
Z^{(c)}_1(\la,k;s) &=& 
\frac{1}{k}\sum_{r\in \bsz_{k}}\left(
 e^{2\pi i \al(\la,r/k)\frac{1}{2}} + e^{-2\pi i \al(\la,r/k)\frac{1}{2}} 
\right) e^{2\pi s\cdot \frac{3}{4}} 
+ \cO(e^{-2\pi s\cdot \frac{1}{4}}) \nn
&=&  \frac{1}{k}\sum_{r\in \bsz_{k}} 
2\left(1-2\sin^2\left(\pi \frac{r}{k}\right)\cos^2(\pi \la)\right)
e^{2\pi s\cdot \frac{3}{4}} 
+ \cO(e^{-2\pi s \cdot \frac{1}{4}}) \nn
&=& 2\sin^2(\pi \la) e^{2\pi s \cdot \frac{3}{4}}
+ \cO(e^{-2\pi s \cdot \frac{1}{4}})~.
\label{evaluation Z c 1}
\end{eqnarray}
This leads us to the tachyonic behavior 
\begin{eqnarray}
Z^{(c)}(\la,k;s) -\frac{1}{s^{(25-p)/2}}
Z^{(c)}_0(\la,k;s)\sim 2 \sin^2(\pi \la)
\frac{1}{s^{(25-p)/2}} 
e^{2\pi s \cdot \frac{3}{4} } ~, ~~ (s\, \rightarrow \, \infty)~,
\label{tachyonic behavior 1}
\end{eqnarray}  
characteristic for the Hagedorn divergence at the very  
high temperature $T= 2T_H$. 
We note that this behavior \eqn{tachyonic behavior 1} 
is universal {\em irrespective of the temperature $T=1/(2\pi k)$.}
Namely, we always face this divergence no matter how low temperature
(including the zero-temperature $k=\infty$) is chosen,  as long as 
the coupling $\la$ is non-zero.
Such effective thermalization as if we were at a very high temperature 
caused by the S-brane has been discussed in \cite{Strominger,MSY} 
based on the minisuperspace approximation, 
and it has been explained as the reflection of 
large rate of open string pair production 
observed by the Unruh detector. 
Our result is likely to be consistent with this observation
at the qualitative level. However, we should note that the production rate 
calculated in \cite{Strominger,MSY} is characterized by $T_H$
rather than $2T_H$.

~

\noindent
{\bf (ii) Near $\la=1/2$ : Effective Hagedorn Behavior of Closed
String Excitations}

As mentioned above, closed string physical states contribute 
to the thermal partition function \eqn{cylinder amplitude 3} 
at the point $\la=1/2$. The thermal instability caused by closed string 
excitations is formally examined in the dual open string channel
despite the absence of physical open string states.
We define   
\begin{eqnarray}
Z(\la,k)&=& \int_0^{\infty} dt \, Z^{(o)}(\la,k;t)~. 
\label{Z o}
\end{eqnarray}
Recalling \eqn{cylinder amplitude 1}, 
we find that $Z^{(o)}(\la,k;t)$ has the following structure
\begin{eqnarray}
Z^{(o)}(\la,k;t)&=& \mbox{const.} \times \frac{1}{t^{1+p/2}}
\sum_{w=0}^{\infty} Z^{(o)}_w(\la,k;t)~,
\label{Z o w}
\end{eqnarray}
where $Z^{(o)}_w(\la,k;t)$ has the form as 
\begin{eqnarray}
Z^{(o)}_w(\la,k;t)& =& \sum_{r\in \bsz_k} \left\lb
e^{2\pi t \left\{1-\left(w+\frac{1}{2}\al(\la,r/k)\right)^2\right\}}
+ e^{2\pi t \left\{1-\left(w-\frac{1}{2}\al(\la,r/k)\right)^2\right\}}
\right\rb \nn
&& ~~~ 
\times \mbox{power series of $e^{-2\pi t}$} ~.
\label{Z o w 2}
\end{eqnarray}
The leading term $Z^{(o)}_0(\la,k;t)$ is again tachyonic 
irrespective of $\la$, $k$, and not relevant for our analysis. 
The next leading term $Z^{(o)}_1(\la,k;t)$ 
is important for the thermal property. We first set $\la=1/2$. 
Recalling \eqn{cylinder amplitude 3}, we find 
$Z^{(o)}_1(1/2,k;t)$ behaves as 
\begin{eqnarray}
Z^{(o)}_1(1/2,k;t)\sim 1 ~, ~~(t\,\rightarrow\,\infty)~,
\end{eqnarray}
since $\al(1/2,z)=0$ holds. 
It means that we have no thermal instability in the closed string 
channel at least for $p\geq 1$.
This fact is quite natural because $\la=1/2$ means that we sit at
a minimum of the tachyon potential from the beginning and 
no particle production seems to be caused.

Now let us consider a small perturbation from the $\la=1/2$ point.
We set $\dsp \la = \frac{1}{2}-\Delta \la$, $0<\Delta \la \ll 1$. 
Based on the simple evaluation
\begin{eqnarray}
\al(1/2-\Delta \la, z) = 2\sin(\pi z) \Delta \la + \cO((\Delta \la)^3)~,
\end{eqnarray}
we can find the tachyonic behavior
\begin{eqnarray}
Z^{(o)}_1(1/2-\Delta \la,k;t)\sim e^{2\pi t \Delta \la'  (2-\Delta \la')}
~, ~~(t\,\rightarrow\,\infty)~,
\end{eqnarray}
where we set $\Delta \la ' \equiv \Delta \la \sin
\left(\frac{\pi}{k}\left\lb\frac{k}{2}\right\rb\right)$ 
($\lb ~~\rb$ is the Gauss symbol).
The IR divergence included in  $Z^{(o)}_1(1/2-\Delta \la,k;t)$
is interpreted in the physical closed string picture 
as the large number  of excitations of {\em all\/} the 
closed string massive modes as if we were above the Hagedorn 
temperature. We again note that this feature does not depend on $k$,
and hence we always face such instability even in 
the zero temperature case. 
It has been shown in \cite{LLM} that we have 
the large emission rate of closed string massive modes
in the brane decay process, which gives  rise to 
a Hagedorn like divergence. Our observation seems to be 
consistent with this fact. 
One should, however, keep it in mind that we have calculated 
the Euclidean cylinder amplitudes,  which has different 
physical meaning compared with that of the Lorentzian signature 
treated in \cite{LLM} (see also \cite{KLMS}) 
in spite of the apparent similarity 
of structure as the BCFT amplitudes.

~


\subsection{Full S-brane in Superstring Case}
\indent

The superstring case is similarly analysed. We will observe the
essentially same thermodynamical behavior, 
but obtain a little modification.

~

\noindent
{\bf (i) Near $\la=0$ : Effective Hagedorn Behavior of 
the Open String Excitations}

We again start with exploring the thermal behavior 
in the open string channel around $\la=0$ point.
We define 
\begin{eqnarray}
Z(\la,k)&=& \int_0^{\infty} dt \, Z^{(c)}(\la,k;s)  ~,\nn
Z^{(c)}(\la,k;s) &=& \mbox{const.}\times \frac{1}{s^{(9-p)/2}} 
\sum_{w=0}^{\infty} Z_w^{(c)}(\la,k;s) ~,
\label{Z c super}
\end{eqnarray}
analogously to \eqn{Z c}. 
The winding $w$ is defined associated to the level 2 theta functions 
$\th_*(is,*)$ (see \eqn{overlap NS}, \eqn{overlap R} 
and \eqn{overlap NS 2}), and $Z_w^{(c)}(\la,k;s)$ has the form 
\begin{eqnarray}
Z_w^{(c)}(\la,k;s) = e^{2\pi s \cdot \frac{1}{2}(1-w^2)}
\times \mbox{power series of $e^{-2\pi s \cdot \frac{1}{2}}$}~.
\end{eqnarray} 
The $w=0$ sector behaves as 
\begin{eqnarray}
Z_0^{(c)}(\la,k;s) \sim 1~,~~(s\,\rightarrow\,\infty)~,
\end{eqnarray}
since the closed string tachyon mode is eliminated by the GSO
projection.
If $\la=0$, since $\al(0,r/(2k))=r/k$, $\beta(0,r/(2k))=r/k$, 
$\gamma(0,(r+1/2)/(2k))=(r+1/2)/k+1/2$
hold, the averaging $\dsp \frac{1}{2k}\sum_{r\in \bsz_{2k}}f(r/k)$
kills all the sectors of $w=1,\ldots, k-1$.
The first non-zero term $Z^{(c)}_{k}(0,k;t)$ behaves as
\begin{eqnarray}
Z_k^{(c)}(0,k;s) &\sim& e^{- 2\pi s \cdot \frac{1}{2}k^2}
\frac{1}{\eta(is)^8}
\left\{\left(\frac{\th_3}{\eta}\right)^4(is) 
+ \left(\frac{\th_4}{\eta}\right)^4(is) \right\}  \nn
&\sim& e^{2\pi s \cdot \frac{1}{2}(1- k^2)}
~,~~(s\,\rightarrow\,\infty)~.
\end{eqnarray}
We so obtain
\begin{eqnarray}
Z^{(c)}(0,k;s) -\frac{1}{s^{(9-p)/2}}Z^{(c)}_0(0,k;s)
\sim \frac{1}{s^{(9-p)/2}}
e^{2\pi s \cdot \frac{1}{2}(1- k^2)} ~, ~~(s\,\rightarrow\,\infty)~,
\end{eqnarray}
reproducing the correct thermal behavior at $T=1/(2\pi\sqrt{2}k)$.
Note that $k=1$  corresponds to  
the Hagedorn temperature for superstring 
(while $T_{k=1}=2T_{H}$ in bosonic string).

Turning on the coupling $\la$, we obtain as in \eqn{evaluation Z c 1}
\begin{eqnarray}
Z^{(c)}_1(\la,k;s) &=&  
\frac{1}{2k} \sum_{r\in \bsz_{2k}}
\left\lb \frac{1}{2}\left\{e^{2\pi i \al(\la, r/(2k))}+
e^{-2\pi i \al(\la, r/(2k))} + e^{2\pi i \beta(\la, r/(2k))}+
e^{-2\pi i \beta(\la, r/(2k))}\right\} \right.  \nn
&& \left. \hspace{1in} 
+ e^{2\pi i \gamma(\la, (r+1/2)/(2k))}+e^{-2\pi i \gamma(\la, (r+1/2)/(2k))}
\right\rb + \cO(e^{-2\pi s \cdot \frac{1}{2}}) \nn
&=& \frac{1}{2k} \sum_{r\in \bsz_{2k}}
\left\lb
-8 \cos^2(\pi \la)+ 8\left\{\sin^4\left(\pi\frac{r}{2k}\right)
+ \cos^4\left(\pi\frac{r}{2k}\right)\right\} \cos^4(\pi\la) \right. \nn
&& \left. \hspace{1in}
+4\sin^2\left(2\pi\frac{r+1/2}{2k}\right)\cos^4(\pi\la)
\right\rb  + \cO(e^{-2\pi s\cdot \frac{1}{2}}) \nn
&=& -4 \sin^2(2\pi \la) + \cO(e^{-2\pi s\cdot \frac{1}{2}})~.
\label{evaluation Z c 1 super} 
\end{eqnarray}
This leads us to the massless IR behavior 
\begin{eqnarray}
Z^{(c)}(\la,k;s) -\frac{1}{s^{(9-p)/2}}
Z^{(c)}_0(\la,k;s)\sim -4  \sin^2(2\pi \la)
\frac{1}{s^{(9-p)/2}} ~, ~~ (s\, \rightarrow \, \infty)~,
\label{massless behavior super}
\end{eqnarray}  
which means that the open string excitations behave 
as if we were at the Hagedorn temperature not depending on $k$.
In this way we have shown the effective thermalization with non-zero 
$\la$ as in the bosonic string case.
We have, however, a ``moderate'' thermal behavior compared with 
the bosonic string. Namely, the behavior \eqn{massless behavior super} 
is massless rather than tachyonic, and hence 
the $t$-integral has no IR divergence as long as $p\leq 6$. 
It is of course originating from the (partial) SUSY cancellation
left even under the thermal boundary condition of world-sheet 
fermions.   


~

\noindent
{\bf (ii) Near $\la=1/2$ : 
Effective Hagedorn Behavior of the Closed String Excitations}

We can again analyse likewise as in the bosonic string.
We write
\begin{eqnarray}
Z(\la,k) =\int_0^{\infty}dt\, Z^{(o)}(\la,k;t)~.
\end{eqnarray} 
At $\la=1/2$, \eqn{cylinder amplitude super 3} 
leads to the massless behavior
\begin{eqnarray}
Z^{(o)}(1/2,k;t) \sim \frac{1}{t^{p/2+1}} ~,~~ (t\,\rightarrow\,\infty)~.
\end{eqnarray}
As is expected, we have no tachyonic divergence, and 
the closed string channel has no thermal instability if $p\geq 1$.

Let us now consider the small perturbation $\la=1/2-\Delta \la$ 
($0<\Delta \la \ll 1$). 
The most important parts included in the thermal partition function
\eqn{cylinder amplitude super 1} are the theta functions such as
$\th_*(is,i\al(\la,*)t) e^{-2\pi t \cdot \frac{1}{2}\al(\la,z)^2}$,
which yields the contribution
\begin{eqnarray}
\sim \frac{1}{2k}\sum_{r\in \bsz_{2k}}\left\lb
e^{-2\pi t \cdot \frac{1}{2}\al(\la, r/(2k))^2} + 
\sum_{w=1}^{\infty} \left\{
e^{-2\pi t\cdot \frac{1}{2}\left(w+\al(\la,r/(2k))\right)^2}
+ e^{-2\pi t\cdot \frac{1}{2}\left(w-\al(\la,r/(2k))\right)^2}
\right\} \right\rb ~,
\end{eqnarray}
and the similar terms including $\beta(\la,*)$ and $\gamma(\la,*)$.
The $w=1$ term is relevant for the desired thermal behavior
and actually gives the leading term. (The thermal boundary condition
for fermions is essential for this fact.) 
Based on the evaluations
\begin{eqnarray}
\al(1/2-\Delta \la, z) &=& 2\sin (\pi z) \Delta \la 
+ \cO\left((\Delta \la)^3\right)~, \nn
\beta(1/2-\Delta \la, z) &=& 1-2 \cos (\pi z) \Delta \la  
+ \cO\left((\Delta \la)^3\right)~, \nn
\gamma(1/2-\Delta \la, z) &=& \frac{1}{2}+\pi \sin (2\pi z) (\Delta \la)^2
+ \cO\left((\Delta \la)^4\right)~,
\end{eqnarray}
we find the tachyonic behavior
\begin{eqnarray}
Z^{(o)}(1/2-\Delta \la,k;t) \sim \frac{1}{t^{1+p/2}} 
e^{4\pi t \Delta \la (1-\Delta \la)}~, ~~(t\,\rightarrow \, \infty)~.
\end{eqnarray}
In this way we have again found the effective thermal instability in the 
closed string channel.


~

\subsection{Half S-branes}
\indent

As we already mentioned, in the half S-brane case
it is natural to rewrite the coupling $\la$ as $\la\equiv \la_0e^{x^0}$
and identify $x^0$ as the real time. We assume that
$\la_0$ is fixed  to be  a positive number of $\cO(1)$, 
and discuss the thermal behaviors in the far past 
$(x^0\, \sim \, -\infty)$  and in the far future 
$(x^0\,\sim\, +\infty)$.

In the far past we should discuss the thermal behavior of on-shell
open string states and the analysis is quite similar to that for the full 
S-brane near the $\la=0$ point. 
In the bosonic string case, 
\eqn{evaluation Z c 1} is replaced with 
\begin{eqnarray}
Z^{(c)}_1(\la_0e^{x^0},k;s) &=& 
\frac{1}{k}\sum_{r\in \bsz_{k}}\left(
 e^{2\pi i \al^{(+)}(\la_0e^{x^0},r/k)\frac{1}{2}} + 
e^{-2\pi i \al^{(+)}(\la_0e^{x^0},r/k)\frac{1}{2}} 
\right) e^{2\pi s\cdot \frac{3}{4}} 
+ \cO(e^{-2\pi s\cdot \frac{1}{4}}) \nn
&=&  \frac{1}{k}\sum_{r\in \bsz_{k}} 
\left\lb (2\pi \la_0e^{x^0})^2 + 2\cos\left(2\pi \frac{r}{k}\right)\right\rb
e^{2\pi s\cdot \frac{3}{4}} 
+ \cO(e^{-2\pi s \cdot \frac{1}{4}}) \nn
&=& (2\pi \la_0e^{x^0})^2 e^{2\pi s\cdot \frac{3}{4}}
+ \cO(e^{-2\pi s \cdot \frac{1}{4}})~,
\label{evaluation Z c 1 half}
\end{eqnarray}
which yields the same tachyonic behavior as in \eqn{evaluation Z c 1}.
On the other hand, in the superstring case, 
we obtain in the similar manner as \eqn{evaluation Z c 1 super}
\begin{eqnarray}
Z^{(c)}_1(\la_0e^{x^0},k;s) &=&  
\frac{1}{2k} \sum_{r\in \bsz_{2k}}\frac{1}{2}
\left\lb e^{2\pi i \al^{(+)}(\la_0e^{x^0}, r/(2k))}+
e^{-2\pi i \al^{(+)}(\la_0e^{x^0}, r/(2k))} \right. \nn
&&  \hspace{.8in}
+ e^{2\pi i \beta^{(+)}(\la_0e^{x^0}, r/(2k))}+
e^{-2\pi i \beta^{(+)}(\la_0e^{x^0}, r/(2k))}    \nn
&&  \hspace{.8in}
+ e^{2\pi i \gamma^{(+)}(\la_0e^{x^0}, (r+1/2)/(2k))}
+e^{-2\pi i \gamma^{(+)}(\la_0e^{x^0}, (r+1/2)/(2k))} \nn
&& \left. \hspace{.8in}
+ e^{2\pi i \gamma^{(+)*}(\la_0e^{x^0}, (r+1/2)/(2k))}
+e^{-2\pi i \gamma^{(+)*}(\la_0e^{x^0}, (r+1/2)/(2k))}
\right\rb 
+ \cO(e^{-2\pi s \cdot \frac{1}{2}}) \nn
&=& \frac{1}{2k} \sum_{r\in \bsz_{2k}} 
\left\lb
4 \cos^2\left(2\pi\frac{r}{2k}\right)
+4 \sin^2\left(2\pi\frac{r+\frac{1}{2}}{2k}\right)-4 \right\rb \nn
&=& 0  + \cO(e^{-2\pi s\cdot \frac{1}{2}})~.
\label{evaluation Z c 1 half super} 
\end{eqnarray} 
We have thus found a massive behavior not depending on $x^0$.
Interestingly, we have the cancellation 
of the massless term in contrast to the full brane case 
\eqn{evaluation Z c 1 super}.

In the far future, the situation is drastically changed. 
We should now examine the thermal behavior of closed string physical 
excitations.
The twist angles $\al^{(+)}(\la_0e^{x^0}, z)$, 
$\beta^{(+)}(\la_0e^{x^0}, z)$, and 
$\gamma^{(+)}(\la_0e^{x^0}, z)$ gain 
very large imaginary parts, explicitly evaluated as
\begin{eqnarray}
\al^{(+)}(\la_0e^{x^0}, z) &\sim& -i\frac{2}{\pi}x^0 ~, \nn
\beta^{(+)}(\la_0e^{x^0}, z) &\sim& 1-i\frac{2}{\pi}x^0 ~, \nn
\gamma^{(+)}(\la_0e^{x^0}, z) &\sim& \frac{1}{2} -i\frac{2}{\pi}x^0 ~.
\end{eqnarray}
We so obtain
\begin{eqnarray}
Z_{\msc{half}}(\la_0e^{x^0},k)
&\approx & k\int_0^{\infty}\frac{dt}{t} \frac{V_p}{(8\pi^2 t)^{p/2}}
\frac{1}{\eta(it)^{24}} \, 
\sum_{n\in \bsz} e^{-\frac{t}{2\pi} 
\left\{2\pi\left(n-\frac{i}{\pi}x^0\right)\right\}^2}~,
\label{cylinder amplitude half future} 
\end{eqnarray}
for bosonic string, and
\begin{eqnarray}
Z_{\msc{half}}(\la_0e^{x^0},k)
&\approx & 2k\int_0^{\infty}\frac{dt}{t} \frac{V_p}{(8\pi^2 t)^{p/2}}
\frac{1}{\eta(it)^{8}} \left(\frac{\th_4}{\eta}\right)^4(it)\, 
\sum_{n\in \bsz} \,
e^{- \frac{t}{2\pi}\left\{2\sqrt{2}\pi
\left(n+\frac{1}{2}-\frac{i}{\pi}x^0\right)\right\}^2}~,
\label{cylinder amplitude half super future} 
\end{eqnarray}
for superstring.
They have quite reminiscent forms of the $\la=1/2$ amplitudes in the
full S-branes \eqn{cylinder amplitude 3} and \eqn{cylinder amplitude
super 3}, which seems consistent with the expectation that 
we only have physical closed string modes after the unstable brane 
collapsed. However, we now have a very strong divergence growing  
exponentially with respect to $(x^0)^2$ \footnote
     {Also in the full S-brane case we can introduce the time dependence
      by replacing the boundary interaction 
   $\dsp \la \int_{\partial \Sigma} d\tau\,\cos X$ with 
   $\dsp \frac{\la_0}{2} \int_{\partial \Sigma}d\tau\, 
   \left(e^{x^0}e^{iX}+e^{-x^0}e^{-iX}\right) \sim 
   \frac{\la_0}{2}\cdot 2\pi i \left(e^{x^0}J^+_0+
  e^{-x^0}J^-_0\right)$. For the far future and past $x^0 \sim \pm
  \infty$,  the twist angles likewise gain large imaginary
  parts, giving rise to the same type divergence.}.
In a naive sense this strong divergence may be
interpreted as the infinite contribution from 
the massive closed string excitations emitted from 
the decaying brane. We will again discuss this point in the last section.

~

\subsection{Comments on Space-like Linear Dilaton Backgrounds}
\indent

Quite recently the S-branes in the general linear dilaton backgrounds  
have been studied in \cite{KLMS} (see also \cite{GS2,MV,KMS,MTV,Schomerus}).
Among other things, the UV divergences in the Lorentzian 
cylinder amplitudes have been shown to be removed 
by considering the space-like linear dilaton characteristic for  
the subcritical string theories. 
Let us briefly discuss whether or not the same mechanism to remove
the divergence works in our cases of thermal amplitudes.

We only focus on the bosonic string case and the superstring case 
is similarly worked out (except for a little modification due to
the GSO projection).
Consider the general conformal system of the form;
\begin{eqnarray}
X \otimes \br_{\phi} \otimes \cM~,
\label{linear dilaton}
\end{eqnarray} 
where $X$ is the Euclidean time coordinate as before and 
$\br_{\phi}$ expresses the CFT with  the linear dilaton 
$\Phi(\phi)=Q\phi$  $(Q \in \br)$. $\cM$ is assumed to be 
an arbitrary unitary CFT. 
The criticality condition now becomes
\begin{eqnarray}
1+\left(1+6Q^2\right) + c_{\cM} =26~.
\label{crit}
\end{eqnarray}
We assume for simplicity that the boundary interaction is 
introduced only along the Euclidean time direction $X$.\footnote
   {In fact, if $Q^2<4$,  we cannot introduce the suitable Liouville
    potential for $\phi$ satisfying the reality condition as is familiar
    in the context of two dimensional gravity.
    In the exceptional case $Q^2=4$ (in other words, with no $\cM$
    sector) we can consider the (boundary) Liouville potential both 
    along the time direction and the $\phi$-direction. This case has been
    intensively studied in the recent papers \cite{MV,KMS,MTV}.}
Namely, we take the world-sheet action \eqn{S E} for the $X$-sector.
The criticality condition leads to the upper bound $Q^2\leq 4$, since 
$c_{\cM}$ must be non-negative.

Now, let us argue on how the thermal behavior is modified 
by the influence of linear dilaton. 
The idea is very simple. In general the cylinder amplitude 
in an arbitrary BCFT behaves as
\begin{eqnarray}
Z(is) \sim e^{-2\pi s \cdot \frac{c_{\msc{eff}}}{24}}~,~~
(s \,\rightarrow\,\infty)~,
\end{eqnarray}
where $s$ is the closed string or open string modulus
(up to some power factor of $s$ in which we are not interested here).
The ``effective central charge'' $c_{\msc{eff}}$ is defined 
in \cite{KutS} (see also \cite{Carlip});
\begin{eqnarray}
c_{\msc{eff}} \equiv c -24 h_0~,
\label{c eff}
\end{eqnarray}
where $h_0$ is the minimal value of conformal weight in the 
spectrum of normalizable states. 
For any unitary CFT we have $c_{\msc{eff}}=c$, since $h_0=0$.
However, in the linear dilaton case we have the mass gap 
$\dsp h_0=\frac{Q^2}{4} >0$ and hence obtain
\begin{eqnarray}
c_{\msc{eff}}= (1+6Q^2)- 24\times \frac{Q^2}{4}=1~.
\end{eqnarray} 
In our previous analysis it amounts to replacing the tachyonic 
factor $e^{2\pi s}$ with an weaker one 
$\dsp e^{2\pi s \left(1-\frac{Q^2}{4}\right)} \equiv 
e^{2\pi s \frac{c_{\cM}}{24}}$.
For example, \eqn{evaluation Z c 1} is now modified as 
\begin{eqnarray}
Z^{(c)}_1(\la,k;s) = 2 \sin^2(\pi \la) e^{2\pi s \cdot 
\left(\frac{3}{4}-\frac{Q^2}{4}\right)} + 
\cO(e^{-2\pi s \cdot \left(\frac{1}{4}+\frac{Q^2}{4}\right)})~.
\end{eqnarray}
Therefore, the effective Hagedorn divergence is removed 
for sufficiently large $Q$ (for sufficiently small $c_{\cM}$, 
in other words). The same mechanism to reduce  the divergence 
works in the other cases. However, since $Q^2$ is at most a number of 
$\cO(1)$, we can never remove the strong divergences 
appearing in \eqn{cylinder amplitude half future} and 
\eqn{cylinder amplitude half super future} 
(the far future amplitude in the half S-brane background describing 
the brane decay).

~

\section{Discussions}
\indent

In this paper we have calculated the thermal partition functions for 
the S-brane backgrounds. We especially examined the thermal property
of the full S-brane case in a detail, and showed that we always have
the Hagedorn like divergence no matter how low temperature is taken,
if the boundary coupling $\la$ is non-zero. 
Parts of such divergences can be removed by considering the space-like
linear dilaton backgrounds, but cannot for the far future amplitude 
in the half S-brane background corresponding to the brane decay. 

The appearance of UV divergences in the thermal 
partition functions is likely to be a signal of the infinite 
contribution to the free energy from the massive closed string 
excitations emitted from the decaying brane. 
However, to be more rigorous it may imply  the failure 
of perturbative calculation in string theory. 
In fact, despite the true marginality and  exact solubility, 
the appearance of divergence in the  modulus integral 
compels us to introduce a cut-off parameter, 
which breaks the conformal invariance.
In string theory the UV divergence in the closed (open) string channel
should be interpreted as the wrong choice of vacuum 
in the dual open (closed) string channel \cite{FS,CLNY2,DR}.
Such ``back-reaction'' problem  was  partly discussed in 
\cite{LLM}, and it was suggested that the $\la=1/2$ point in 
the full S-brane may be the solution correctly incorporating 
the back-reaction. It might be amazing and suggestive that our 
divergent amplitudes of half S-brane in the far future 
\eqn{cylinder amplitude half future} and 
\eqn{cylinder amplitude half super future} have the reminiscent forms 
of the $\la=1/2$ amplitudes \eqn{cylinder amplitude 3} 
and \eqn{cylinder amplitude super 3}. We also point out that 
the $\la=1/2$ superstring amplitude \eqn{cylinder amplitude super 3}
is actually finite (for generic choice of $p$), suggesting 
the correctly chosen vacuum. 
In any case, the appearance of Hagedorn like divergence 
seems to suggest that the stringy correction 
is not small and the back-reaction is not negligible 
{\em even under the weak coupling limit $g_s \,\rightarrow\, 0$}. 
(In fact, we observed the divergence 
in the thermal partition function as free string theory.)
To study the back-reaction seriously is surely
the most important and challenging task and we deserve it for future study.

It is an also interesting subject to perform the similar thermal analysis 
in the general linear dilaton backgrounds. Especially, it is a non-trivial 
problem to ask whether our calculation of thermal amplitudes can be
extended to the S-brane backgrounds with the non-vanishing  time component 
of linear dilaton. 
Presumably, the ``boundary minimal models'' would play important roles,
and the quantum $SU(2)$ algebra might provide useful tools of calculation.

~

~


\section*{Acknowledgements}
\indent

I would like to thank S. Sugimoto and E. Watanabe for valuable discussions.
I am also grateful to Y. Matsuo for his nice lecture about the paper 
\cite{LLM} at Univ. of Tokyo, which inspired me to start this work.

This research is partially supported by 
a Grant-in-Aid for the Encouragement of Young Scientists 
($\sharp 15740138$)  from 
Japanese Ministry of Education, 
Culture, Sports, Science and Technology.

\newpage

\section*{Appendix ~ Some Notations}
\setcounter{equation}{0}
\def\theequation{A.\arabic{equation}}

We here summarize the convention of theta functions.
We set $q\equiv e^{2\pi i \tau}$, $y\equiv e^{2\pi i z}$.
\begin{eqnarray}
&& \th_1(\tau,z) =i\sum_{n=-\infty}^{\infty}(-1)^n q^{(n-1/2)^2/2} y^{n-1/2}
  \equiv 2 \sin(\pi z)q^{1/8}\prod_{m=1}^{\infty}
    (1-q^m)(1-yq^m)(1-y^{-1}q^m)~, \nn
&&  \th_2(\tau,z)=\sum_{n=-\infty}^{\infty} q^{(n-1/2)^2/2} y^{n-1/2}
  \equiv 2 \cos(\pi z)q^{1/8}\prod_{m=1}^{\infty}
    (1-q^m)(1+yq^m)(1+y^{-1}q^m)~, \nn
&& \th_3(\tau,z)=\sum_{n=-\infty}^{\infty} q^{n^2/2} y^{n}
  \equiv \prod_{m=1}^{\infty}
    (1-q^m)(1+yq^{m-1/2})(1+y^{-1}q^{m-1/2})~, \nn
&& \th_4(\tau,z)=\sum_{n=-\infty}^{\infty}(-1)^n q^{n^2/2} y^{n}
  \equiv \prod_{m=1}^{\infty}
    (1-q^m)(1-yq^{m-1/2})(1-y^{-1}q^{m-1/2})~ .
\end{eqnarray}
 \begin{eqnarray}
 \Th{m}{k}(\tau,z)&=&\sum_{n=-\infty}^{\infty}
 q^{k(n+\frac{m}{2k})^2}y^{k(n+\frac{m}{2k})} , 
 \end{eqnarray}
 We often use the abbreviations; $\th_i \equiv \th_i(\tau, 0)$
 ($\th_1\equiv 0$), $\Th{m}{k}(\tau) \equiv \Th{m}{k}(\tau,0)$.
 We also use the standard convention of $\eta$-function;
 \begin{equation}
 \eta(\tau)=q^{1/24}\prod_{n=1}^{\infty}(1-q^n)~.
 \end{equation}
The affine character of $SU(2)_k$ with spin $\ell/2$ ($0\leq \ell \leq k$)
is given by the formula
\begin{equation}
\chi^{(k)}_{\ell}(\tau, z) =\frac{\Th{\ell+1}{k+2}(\tau,z)
-\Th{-\ell-1}{k+2}(\tau,z)}
{\Th{1}{2}(\tau,z)-\Th{-1}{2}(\tau,z)}~.
\end{equation}

~

\newpage

\end{document}